\documentclass[prd,aps,preprint,groupedaddress,superscriptaddress,nofootinbib]{revtex4}
\usepackage{graphicx}
\usepackage{amsmath,amssymb,bm,mathrsfs}
\usepackage{hyperref}
\usepackage{autobreak,mathtools}
\usepackage{comment}
\usepackage[dvipsnames, usenames]{xcolor}
\usepackage[scr=boondoxo,scrscaled=1.05]{mathalfa}
\usepackage{subfig}
\usepackage{ulem}

\definecolor{rRGB}{RGB}{171, 40, 52}

\allowdisplaybreaks[1]

\begin{document}

\title{Effective geometrostatics of spherical stars\\ beyond general relativity}
\author{Julio Arrechea}
\affiliation{SISSA - International School for Advanced Studies, Via Bonomea 265, 34136 Trieste, Italy}
\affiliation{
IFPU - Institute for Fundamental Physics of the Universe, Via Beirut 2, 34014 Trieste, Italy}
\affiliation{INFN Sezione di Trieste, Via Valerio 2, 34127 Trieste, Italy}
\author{Ra\'ul Carballo-Rubio}
\affiliation{Instituto de Astrof\'isica de Andaluc\'ia (IAA-CSIC),
Glorieta de la Astronom\'ia, 18008 Granada, Spain}
\author{Matt Visser}
\affiliation{
School of Mathematics and Statistics, Victoria University of Wellington, PO Box 600, Wellington 6140, New Zealand
}
\begin{abstract}
We provide a set of general tools to study the problem of stellar equilibrium in any gravitational theory in which spherically symmetric spacetimes satisfy master field equations taking the form of an equality between an identically conserved tensor, with derivatives of up to second order in the metric, and an identically conserved matter tensor. We derive the most general expression for the Tolman--Oppenheimer--Volkoff equation of stellar equilibrium that is compatible with these minimal requirements. A general discussion of the conditions that guarantee geodesic completeness at the center of symmetry is also presented. The equations of stellar equilibrium are integrated in a subset of the space of allowed deformations of general relativity proposed by Ziprick and Kunstatter, allowing us to illustrate universal aspects associated with the weakening of the strength of gravity, such as the mitigation of the Buchdahl limit obtained in general relativity or the existence of static solutions describing regular black holes with perfect fluid cores.

\bigskip


\bigskip

{\sc Dated:} Tuesday 17 March 2026; \LaTeX-ed \today.

\end{abstract}

\maketitle

\tableofcontents

\bigskip
\bigskip

\hrule\hrule\hrule

\clearpage

\section{Introduction}

The search for alternative theories of gravity has been the focus of numerous research works and programmes. The motivations underlying such a search are diverse, from the more pragmatic to the more ideological. On the pragmatic side, developing alternative theories to general relativity is an important step in the deployment of precision tests of Einstein's theory~\cite{Will:2014kxa,Berti:2015itd,Wolf:2023xrv,Ayzenberg:2023hfw,Yunes:2024lzm,Gupta:2025utd}. From a more ideological perspective, finding theories that combine the principles of quantum mechanics with the geometric description of gravity characteristic of general relativity is expected to provide a deeper understanding of the fabric of our universe~\cite{Rovelli:2000aw,Carlip:2001wq,Rovelli:2004tv,Stachel:2005ij,Loll:2022ibq,deBoer:2022zka,Buoninfante:2024yth,Crowther:2025cfx}.

Black holes have been receiving increasing attention due to playing an important role in several of these approaches,  and being a convergence point for motivations of diverse nature. Being a paradigmatic example of a gravitational system displaying strong gravity effects puts them at the center of many approaches aiming at testing general relativity~\cite{Johannsen:2015mdd,Yagi:2016jml,Eckart:2017bhq,Carballo-Rubio:2018jzw,Cardoso:2019rvt,Buoninfante:2024oxl,Visser:2008rtf}. At the same time, the incompleteness of black hole interiors in general relativity is expected to be remediable only in theories beyond general relativity, thus signaling the need for new physics in order to achieve a complete description of these gravitational systems~\cite{Perez:2017cmj,Eichhorn:2022bgu,Carballo-Rubio:2025fnc}. 

It is therefore common to consider black hole physics as a testing ground for the development of alternative theories to general relativity. Doing so allows trying out proposals for the behavior of gravity in extreme conditions in simpler settings in which symmetries (spherical symmetry or axisymmetry) are present, and in which the matter content can be taken as trivial (vacuum) in a first instance, as the gravitational structure of a black hole does not require the presence of matter fields for its definition. Once the modified behavior of gravity in black hole interiors resulting from one of these proposals is understood and verified to satisfy some consistency criteria, it may then be interesting to study the implications of the same underlying principles to other situations and determine any resulting deviations with respect to general relativity.

In this paper, we will be working with a formalism that relies mostly on symmetry arguments and that follows a bottom-up approach, similar to effective field theory in particle physics, for the study of gravitational systems with symmetries. In practice, this approach involves using lower-dimensional field theories to describe the dynamics of higher-dimensional spacetimes with symmetries. Ziprick and Kunstatter~\cite{Ziprick:2010vb} first developed simulations of the formation of spherically symmetric regular black holes using a 2-dimensional dilaton gravity as an effective theory. This phenomenological 2-dimensional approach was then further developed in references~\cite{Taves:2014laa,Kunstatter:2015vxa} (see related numerical explorations in recent works~\cite{Barenboim:2024dko,Barenboim:2025ckx}). More recently, a set of explicitly 4-dimensional field equations has been constructed by coupling 2-dimensional Horndeski theory to additional matter fields, providing the most general form of the field equations for spherically symmetric gravitational fields in which the Einstein tensor is deformed into a conserved tensor constructed from up to second-order derivatives of the metric~\cite{Carballo-Rubio:2025ntd}. Here, we will be exploring the implications of these equations for the problem of stellar equilibrium.

This paper is organized as follows. We review the formalism and field equations to be used in the rest of the paper in Sec.~\ref{Sec:SphSym}. The general equations of stellar equilibrium in this setup are derived in Sec.~\ref{Sec:SteEqu}, together with a discussion of conditions that guarantee geodesic completeness. The properties of solutions of the equations of stellar equilibrium for specific subfamilies of theories are presented in Sec.~\ref{Sec:Sols}. Some final remarks and considerations are collected in Sec.~\ref{sec:Conclusions}.

\section{Effective geometrodynamics of spherical spacetimes}
\label{Sec:SphSym}

In this section we shall set up and analyze the most general second-order differential equations relevant to describing the dynamics of spherically symmetric spacetimes~\cite{Boyanov:2025pes,Carballo-Rubio:2025ntd}.

\subsection{Master field equations}

We will be working with spherically symmetric spacetimes presented in so-called ``warped product'' form~\cite{semi-riemannian}:
\begin{equation}
g_{\mu\nu}(y)\;\text{d}y^\mu\text{d}y^\nu=
q_{ab}(x)\;\text{d}x^a\text{d}x^b
+r^2(x)\;\gamma_{ij}(\theta)\;\text{d}\theta^i\text{d}\theta^j.
\end{equation}
Here $\{y^\mu\}_{\mu=0}^{D-1}$ are $D$--dimensional coordinates, whereas $q_{ab}(x)$ and $r(x)$ are the 2-metric and a scalar field, respectively, on the 2-dimensional radial-temporal spacetime equipped with  coordinates $\{x^a=x^0,x^1\}$. The function $r(x)$ is called the ``warping factor'', since setting $r(x)\to 1$ results in the considerably simpler ``direct product'' spacetime.  
Furthermore the  $\{\theta^i\}_{i=1}^{D-2}$ are angular coordinates, and $\text{d}\Omega_{(D-2)}^2=\gamma_{ij}(\theta) \; \text{d}\theta^i\text{d}\theta^j$ is the line element on the unit $(D-2)$--sphere. 
Explicitly, $\text{d}\Omega_{(D-2)}^2= \sum_{i=1}^{D-2} [\Pi_{j=1}^{i-1} \sin^2(\theta_j)] \; d\theta_i^2$.

In matrix form we can write this as
\begin{equation}
    g_{\mu\nu}(y) = \left[\begin{array}{c|c} q_{ab}(x) & 0 \\ \hline 0 & r^2(x)\;\gamma_{ij}(\theta) \end{array}\right].
\end{equation}

The most general field equations governing dynamics of spherically symmetric spacetimes in which the Einstein tensor is deformed into an identically conserved tensor containing up to second-order derivatives of the fields $q_{ab}(x)$ and $r(x)$ can be written as~\cite{Carballo-Rubio:2025ntd}
\begin{equation}\label{eq:feqs}
\mathscr{G}_{\mu\nu}(q,r)=8\pi T_{\mu\nu},
\end{equation}
where
\begin{equation}\label{eq:mfeqs}
\mathscr{G}_{\mu\nu}(q,r)=
{r^{2-D}}\mathscr{E}_{ab}(q,r)\;\delta^a_\mu\; \delta^b_\nu
-\frac{r^{5-D}}{2(D-2)}\mathscr{F}(q,r)\; \gamma_{ij}(\theta)\; \delta^i_\mu\; \delta^j_\nu,
\end{equation}
that is
\begin{equation}
    \mathscr{G}_{\mu\nu}(q,r)=
 \left[\begin{array}{c|c} r^{2-D}\mathscr{E}_{ab}(q,r) & 0 \\ \hline 
 0 & -\frac{r^{5-D}}{2(D-2)}\mathscr{F}(q,r)\; \gamma_{ij}(\theta) \end{array}\right].
\end{equation}
Here $\mathscr{E}_{ab}(q,r)$ and $\mathscr{F}(q,r)$ are obtained as the variations, with respect to $q^{ab}(x)$ and $r(x)$ respectively, of the 2\nobreakdash-dimensional Horndeski action~\cite{Horndeski:1974wa,Kobayashi:2019hrl}:
\begin{equation}\label{eq:horact}
\mathcal{L}_{\rm 2DH}=H_2(r,\chi)-H_3(r,\chi)\square r+H_4(r,\chi)\mathcal{R}-2\partial_\chi H_{4}(r,\chi)\left[(\square r)^2-\nabla^a\nabla^b r\nabla_a\nabla_b r\right],
\end{equation}
where $\mathcal{R}$ is the Ricci scalar of a 2-dimensional metric $q_{ab}$, $\{H_i(r,\chi)\}_{i=2}^4$ are generic functions of two variables, and the convention $\chi=\left(\nabla r\right)^2$ is used. Specifically:
\begin{align}\label{eq:geneqs1}
\mathscr{E}_{ab}(q,r)=\frac{1}{\sqrt{-q}}\frac{\delta\mathcal{L}}{\delta q^{ab}}&=\bm{\beta}\nabla_a\nabla_b r-q_{ab}\left(\frac{1}{2}\bm{\alpha}+\bm{\beta}\square r \right)+\left(\partial_\chi\bm{\alpha}-\partial_r\bm{\beta} \right)\nabla_ar\nabla_br
,\\
\mathscr{F}(q,r)=\frac{1}{\sqrt{-q}}\frac{\delta\mathcal{L}}{\delta r}&=-\bm{\beta}\mathcal{R}+2\partial_r\bm{\beta}\square r+\partial_r\bm{\alpha}+2\partial_\chi\bm{\beta}\left[\left(\square r\right)^2-\nabla_a\nabla_b r\nabla^a\nabla^br\right]\nonumber\\
&-2\partial_r\left(\partial_\chi\bm{\alpha}- \partial_r\bm{\beta}\right)\chi-2\left(\partial_\chi\bm{\alpha}-\partial_r\bm{\beta} \right)\square r-2\partial_\chi\left(\partial_\chi\bm{\alpha}-\partial_r\bm{\beta}\right)\nabla_a r\nabla^a\chi.
\label{eq:geneqs2}
\end{align}
Here $\bm{\alpha}(r,\chi)=H_2+\chi\partial_r\left(H_3-2\partial_r H_4\right)$ and $\bm{\beta}(r,\chi)=\chi\partial_\chi\left(H_3-2\partial_r H_4\right)-\partial_rH_4$ are externally specified and essentially arbitrary functions of $r$ and $\chi$ (that only two functions arise in the field equations indicates that the Lagrangian in Eq.~\eqref{eq:horact} is overparametrized, which is a conscious choice to simplify making contact with the literature on Horndeski theory). Note that both $\bm{\alpha}[r(x),\chi(x)]$ and $\bm{\beta}[r(x),\chi(x)]$ carry both explicit dependence on $r$ and $\chi$, and implicit dependence on the coordinates $x$, which becomes an implicit dependence on $r(x)$ whenever the latter scalar field is used as one of the coordinates.

These two functions control all the features of the gravitational interaction in these theories. In particular, we will see more explicitly that, for the static solutions studied in this paper, $\bm{\beta}$ can be directly related to the strength of gravity. While in this paper we will be working explicitly with $\bm{\alpha}$ and $\bm{\beta}$, it is straightforward to calculate the functions $\{H_i\}_{i=2}^4$ for the different examples below by using the relations $H_2=\bm{\alpha}$, $H_3=-2\bm{\beta}$, $H_4=-\int\text{d}r\,\bm{\beta}$~\cite{Boyanov:2025pes}.

Two specific special cases are of particular interest:
\begin{itemize}
\item 
If $\partial_\chi\bm{\alpha}=\partial_r\bm{\beta}$, implying the existence of a potential function $\bm{\Omega}(r,\chi)$ satisfying both $\bm{\alpha}=\partial_r \bm{\Omega}$ and $\bm{\beta}=\partial_\chi\bm{\Omega}$, then we have 
a significant number of simplifications:
\begin{align}\label{eq:poteqs1}
\mathscr{E}_{ab}(q,r)=\frac{1}{\sqrt{-q}}\frac{\delta\mathcal{L}}{\delta q^{ab}}&=\partial_\chi\bm{\Omega}\nabla_a\nabla_b r-q_{ab}\left(\frac{1}{2}\partial_r \bm{\Omega}+\partial_\chi\bm{\Omega}\square r \right)
,\\
\mathscr{F}(q,r)=\frac{1}{\sqrt{-q}}\frac{\delta\mathcal{L}}{\delta r}&=-\partial_\chi\bm{\Omega}\mathcal{R}+\partial_r\partial_\chi\bm{\Omega}\square r+\partial_r^2\bm{\Omega}+2\partial^2_\chi\bm{\Omega}\left[\left(\square r\right)^2-\nabla_a\nabla_b r\,\nabla^a\nabla^br\right].
\label{eq:poteqs2}
\end{align}
\item
The spherically symmetric Einstein field equations are recovered as the particular case
\begin{equation}\label{eq:grchoicesalphabetaD}
\bm{\alpha}_{\rm GR}=(D-2)(D-3)r^{D-4}(1-\chi),\quad
\bm{\beta}_{\rm GR}=-(D-2)r^{D-3},\quad
\bm{\Omega}_{\rm GR}= (D-2)r^{D-3} (1-\chi),
\end{equation}
for which Eqs.~(\ref{eq:geneqs1}-\ref{eq:geneqs2}) and even Eqs.~(\ref{eq:poteqs1}-\ref{eq:poteqs2}) are greatly simplified. In particular, the last term in the first line of Eq.~\eqref{eq:geneqs2} vanishes because $\partial_\chi\bm{\beta}_{\rm GR}=0$, while all the terms in the second line of Eq.~\eqref{eq:geneqs2} vanish due to the fact that $\partial_\chi\bm{\alpha}_{\rm GR}=\partial_r\bm{\beta}_{\rm GR}$. That is, for $D$-dimensional general relativity one has
\begin{eqnarray}\label{eq:GReqs1}
\mathscr{E}_{ab}(q,r)&=&\frac{1}{\sqrt{-q}}\frac{\delta\mathcal{L}}{\delta q^{ab}}
=-(D-2)\bigg\{ r^{D-3} \nabla_a\nabla_b r\nonumber\\
&&\qquad\qquad\qquad -q_{ab}\left(\frac{1}{2}
(D-3)r^{D-4} (1-\chi) -r^{D-3} \square r \right)\bigg\}
,\\
\mathscr{F}(q,r)&=&\frac{1}{\sqrt{-q}}\frac{\delta\mathcal{L}}{\delta r}
=(D-2) \bigg\{ r^{D-3}\mathcal{R}-(D-3) r^{D-4}\square r \nonumber\\
&&\qquad\qquad\qquad -(D-3)(D-4) r^{D-5}(1-\chi)\bigg\}.
\label{eq:GReqs2}
\end{eqnarray}
Unsurprisingly, general relativity completely trivializes in $D=2$ dimensions, and almost trivializes in $D=3$ dimensions. Even $D=4$ general relativity is somewhat simpler than the higher dimensional case.
\end{itemize}

On the other hand, the most general matter source compatible with these symmetries is 
\begin{equation}\label{eq:Ddimsource}
T^{(D)}_{\mu\nu}=T_{ab}\; \delta^a_\mu\delta^b_\nu+p_t\;r^2\;\gamma_{ij}\;\delta^i_\mu\delta^j_\nu,
\end{equation}
that is
\begin{equation}
    T^{(D)}_{\mu\nu} = \left[\begin{array}{c|c} T_{ab} & 0 \\ \hline 0 & p_t\; r^2(x)\;\gamma_{ij}(\theta) \end{array}\right].
\end{equation}
Here $T_{ab}$ contains information about the energy density, radial pressure and fluxes in the 2-dimensional sector, while $p_t$ is the tangential pressure. 

\section{Stellar equilibrium}
\label{Sec:SteEqu}

In this section we shall derive a generalization of the equations of stellar equilibrium, in particular the Tolman--Oppenheimer--Volkoff equation, and analyze the conditions that guarantee geodesic completeness at the center of symmetry $r=0$.

\subsection{Equilibrium equations} 

As the starting point in the analysis of the problem of stellar equilibrium, we consider a geometric ansatz accommodating the spacetimes of interest. Aside from these spacetimes being static, using the scalar field $r(x)$ as one of the coordinates simplifies the resulting equations. The static 2--dimensional line element is then written as
\begin{equation}\label{eq:2gstaticchi}
q_{ab}(x)\;\text{d}x^a\text{d}x^b=-e^{2\nu(r)}\; \text{d}t^2+\frac{\text{d}r^2}{\chi(r)}.
\end{equation}
In order to make contact with the existing literature, we can rewrite the function $\chi(r)$ in terms of the $D$-dimensional Misner--Sharp quasilocal mass,
\begin{equation}\label{eq:chionshell}
\chi(r)
=1-\frac{2m(r)}{r^{D-3}},
\end{equation}
so that Eq.~\eqref{eq:2gstaticchi} becomes
\begin{equation}\label{eq:2gstatic}
q_{ab}(x)\;\text{d}x^a\text{d}x^b=-e^{2\nu(r)}\; \text{d}t^2+\frac{\text{d}r^2}{1-2m(r)/r^{D-3}}.
\end{equation}
The only non-zero Christoffel symbols for the 2-metric $q_{ab}$ are then
\begin{equation}\label{eq:christoffel}
\Gamma^{t}_{tr}=\nu',\quad \Gamma^r_{tt}=e^{2\nu}\left(1-2m/r^{D-3}\right)\nu',\quad \Gamma^r_{rr}=-\frac{(D-3)m/r^{D-2}-m'/r^{D-3}}{1-2m/r^{D-3}}.
\end{equation}
Further useful relations are
\begin{equation}\label{eq:relations}
\nabla_t\nabla_t r=-\Gamma^r_{tt},\quad \nabla_r\nabla_r r=-\Gamma^r_{rr},\quad \square r=-e^{-2\nu}\Gamma^r_{tt}-\left(1-2m/r^{D-3}\right)\Gamma^r_{rr}.
\end{equation}
On the other hand, the matter source~\eqref{eq:Ddimsource} has components
\begin{equation}\label{eq:pfluid}
T_{tt}=\rho \; q_{tt}, \quad T_{tr}=0,\quad T_{rr}=p_r\; q_{rr},\quad T_{ij}=p_t\; g_{ij}.
\end{equation}

The following two differential equations are obtained from Eq.~\eqref{eq:feqs}:
\begin{align}
2(D-3)\bm{\beta} r^{2-D}m+\bm{\alpha}-2\bm{\beta} r^{3-D}m' &=16\pi r^{D-2}\rho,\label{eq:tteq}\\
-\bm{\beta}\nu'-\frac{\bm{\alpha}}{2\left(1-2m/r^{D-3}\right)}+\partial_\chi\bm{\alpha}-\partial_r\bm{\beta}&=\frac{8\pi r^{D-2}p_r}{1-2m/r^{D-3}}.\label{eq:rreq}
\end{align}
Note that these equations are valid for generic functions $\bm{\alpha}(r,\chi)$ and $\bm{\beta}(r,\chi)$. To be more precise about this, observe that $\bm{\alpha}(r,\chi) = \bm{\alpha}(r,1-2m/r^{D-3}) =\bm{\tilde\alpha}(r,m) \to \bm{\tilde\alpha}(r,m(r))$.  Likewise $\bm{\beta}(r,\chi) = \bm{\beta}(r,1-2m/r^{D-3}) =\bm{\tilde\beta}(r,m) \to \bm{\tilde\beta}(r,m(r)) $, and similarly the functional derivatives $\partial_\chi\bm{\alpha}$ and $\partial_r\bm{\beta}$ exhibit the same behaviour. Keeping this in mind, for brevity we often suppress the explicit arguments and implied functional form for the functions  $\bm{\alpha}$,  $\bm{\beta}$, and their derivatives.

Instead of explicitly writing  the angular equation (\ref{eq:geneqs2}), the equation proportional to $\mathscr{F}(q,r)$, we can use the conservation equation for the fluid:
\begin{equation}\label{eq:cons}
2(p_r-p_t)+rp'_r+r\left(\rho+p_r\right)\nu'=0.
\end{equation}
We thus have a closed system of equations, Eqs.~(\ref{eq:tteq}--\ref{eq:cons}), for $m(r)$, $p(r)$, and $\nu(r)$ --- once $\bm{\alpha}(r,\chi)$ and $\bm{\beta}(r,\chi)$ are specified as functions of $r$ and $\chi=1-2m/r^{D-3}$.

This system of equations satisfies several remarkable properties. The first one is that it can be solved as a sequence of first-order ordinary differential equations for each of the unknown functions, as is the case in general relativity. Indeed, Eq.~\eqref{eq:tteq} is a first-order ordinary differential equation for $m(r)$, which we can rearrange as
\begin{equation}\label{eq:genmass}
\bm{\alpha}-2\bm{\beta}\frac{\text{d}}{\text{d}r}\left(m/r^{D-3}\right) =16\pi r^{D-2}\rho.
\end{equation}
That is, for once being totally explicit regarding all functional dependencies, we wish to solve the first-order nonlinear ordinary differential equation
\begin{equation}\label{eq:genmass2}
\bm{\tilde\alpha}(r,m(r))-2\bm{\tilde\beta}(r, m(r))\;
\frac{\text{d}}{\text{d}r}\left(m(r)/r^{D-3}\right) =16\pi r^{D-2}\rho(r).
\end{equation}
Once $m(r)$ is determined, we can combine Eqs.~\eqref{eq:rreq} and~\eqref{eq:cons} to yield the generalized Tolman--Oppenheimer--Volkoff (TOV) differential equation, relevant to any spherically symmetric body in any extension of general relativity in which the Einstein tensor is deformed into an identically conserved tensor constructed solely from up to second derivatives of the metric:
\begin{equation}\label{eq:gentov}
p'_r=\frac{\rho+p_r}{2\bm{\beta}}\left[\frac{16\pi r^{D-2}p_r
+\bm{\alpha}}{1-2m/r^{D-3}}-2\left(\partial_\chi\bm{\alpha}-\partial_r\bm{\beta}\right)\right]-\frac{2\left(p_r-p_t\right)}{r}.    
\end{equation}
This differential equation can be solved to obtain $p(r)$ for specific theories (or families of theories). The remaining metric function $\nu(r)$ is then obtained by solving Eq.~\eqref{eq:cons}.

It is remarkable that the minimal assumptions on the symmetries of the problem and the differential order of the equations of motion, which are behind the construction of the master field equations in Eq.~\eqref{eq:feqs}, also result in such a simple generalization of this well-known TOV equation describing stellar equilibrium in general relativity.

It is straightforward to check that Eqs.~(\ref{eq:genmass}-\ref{eq:gentov}) include $D$-dimensional general relativity as a particular case~\cite{PoncedeLeon:2000pj}, which can be shown using Eqs.~\eqref{eq:grchoicesalphabetaD} and~\eqref{eq:chionshell} to write
\begin{equation}
\bm{\alpha}_{\rm GR}=(D-2)(D-3)\frac{2m}{r}
=-2(D-3) \bm{\beta}_{\rm GR}  r^{2-D}m,\quad 
\bm{\beta}_{\rm GR}=-(D-2)r^{D-3}.
\end{equation}
Whence in $D$-dimensional general relativity the (anisotropic) TOV becomes
\begin{equation}
m' = {8\pi\over D-2} \; \rho\; r^{D-2};   
\qquad
p'_r = - {\rho+p_r\over r^{D-2} } \; { (D-3) m + 8\pi r^{D-1}p_r/\left(D-2\right) \over 1 - 2m /r^{D-3} }
-{2(p_r-p_t)\over r}.
\end{equation}
The anisotropic part of this equation is generic to arbitrary dimensionality. 
The isotropic part of the $D$-dimensional TOV can be checked against many published references, among which we mention~\cite{Bagchi:2020sja, Bakhti:2020yue, PoncedeLeon:2000pj,Bueno:2025tli}.

In particular, for $D=4$ have
\begin{equation}
\bm{\alpha}_{\rm GR}={4m\over r}; \qquad \bm{\beta}_{\rm GR} = -2r; 
\end{equation}
and so we recover the well-known expressions~\cite{Oppenheimer:1939ne}
\begin{equation}
m' =4\pi r^{2}\rho,\qquad p'_r=-\frac{\rho+p_r}{r^2}\left(\frac{m + 4\pi p_r \; r^{{3}} }{1-2m/r}\right)-\frac{2\left(p_r-p_t\right)}{r}.
\end{equation}
%

\subsection{Geodesic completeness}\label{sec:Parity}

Let us now formulate some general observations on the sort of consistency conditions one might wish to impose in order to guarantee that the lower-dimensional fields $q_{ab}(x)$ and $r(x)$ can be used to reconstruct a $D$-dimensional metric~\cite{Carballo-Rubio:2025ntd}.
Appealing to ``local flatness'' of the manifold, any sufficiently small ball (of coordinate radius $r_0$ say)  around the origin should locally approximate a ball in flat Minkowski space to arbitrary accuracy.
In particular, this means that pressure, density, and other local observables should be arbitrarily smooth --- mathematically that can be phrased as being $C^\infty$ on the entire ball.
In contrast, quasi-local observables, such as the Misner--Sharp quasi-local mass need not be $C^\infty$ on the entire ball, but it would still be a good idea to impose $C^\infty$ behaviour on $r \in (0,r_0)$.

Now mathematically ``analyticity'' ($C^\omega$) is stronger than ``infinitely smooth'' ($C^\infty$), the distinction being that analyticity requires the Taylor series to have a non-zero radius of convergence.
For example:
\begin{itemize}
\item 
For $a> 0$ the function $e^{-a/r}$ is $C^\infty$ on $(0,\infty)$,  but is not $C^\omega$ at $r=0$. 
\item
For $a\neq  0$ the function $e^{-a^2/r^2}$ is $C^\infty$ on $(-\infty,\infty)$,  but is not $C^\omega$ at $r=0$. 
\item
Many generalizations possible.
\end{itemize}
Smooth functions of this type were used in developing various ``Minkowski core'' RBHs~\cite{Simpson:2019mud,Simpson:2021dyo, Simpson:2021zfl}.

In the following, we will assume analyticity and, therefore, the existence of a Taylor series expansion. If physical observables (density, pressure, etc...) are analytic then they must be described by even parity Taylor series:
\begin{equation}
p(r) = \sum_{n=0}^\infty {p^{(2n)}|_{r=0}\over(2n)!} \;r^{2n}; \qquad
\rho(r) = \sum_{n=0}^\infty {\rho^{(2n)}|_{r=0}\over(2n)!} \; r^{2n}. \qquad
\end{equation}
This even parity property must be true regardless of dimensionality, and regardless of the dynamics.
Note that once one goes above the surface of the star $p\to 0$ and $\rho\to 0$, and zero is an even function.
We can construct a direct and explicit proof of this statement by demanding invariance of scalar functions with respect to the Killing vectors that generate rotations:
\begin{equation}
\xi_x=y\partial_z-z\partial_y,\quad \xi_y=z\partial_x-x\partial_z,\quad \xi_z=x\partial_y-y\partial_x.    
\end{equation}
For a scalar function $\psi(t,x,y,z)$ that is invariant under rotation, we should have
\begin{equation}
\mathcal{L}_{\xi_x}\psi=\mathcal{L}_{\xi_y}\psi=\mathcal{L}_{\xi_z}\psi=0.    
\end{equation}
Let us take the first equation, for instance. We then have:
\begin{equation}
\mathcal{L}_{\xi_x}\psi=y\partial_z\psi-z\partial_y\psi=0.    
\end{equation}
The derivatives of the expression above in terms of $\psi$ and its first-order derivatives must also vanish, which allows us to write:
\begin{align}
\partial_y\mathcal{L}_{\xi_x}\psi&=y\partial_y\partial_z\psi-z\partial_y^2\psi+\partial_z\psi=0,\nonumber\\
\partial_z\mathcal{L}_{\xi_x}\psi&=y\partial_z^2\psi-z\partial_y\partial_z\psi-\partial_y\psi=0,\nonumber\\
\partial_y^2\mathcal{L}_{\xi_x}\psi&=y\partial_y^2\partial_z\psi-z\partial_y^3\psi+2\partial_y\partial_z\psi=0,\nonumber\\
\partial_z\partial_y\mathcal{L}_{\xi_x}\psi&=y\partial_y\partial_z^2\psi-z\partial_y^2\partial_z\psi+\partial_z^2\psi-\partial_y^2\psi=0,\nonumber\\
\partial_z^2\mathcal{L}_{\xi_x}\psi&=y\partial_z^3\psi-z\partial_y\partial_z^2\psi-2\partial_y\partial_z\psi=0.
\end{align}
This procedure can be implemented up to any arbitrary order in the derivatives.

Evaluating these at $x=y=z=0$, we obtain
\begin{equation}
\left.\partial_y\psi\right|_{x=y=z=0}=\left.\partial_z\psi\right|_{x=y=z=0}=\left.\partial_y\partial_z\psi=0\right|_{x=y=z=0},\quad \left.\partial_y^2\psi\right|_{x=y=z=0}=\left.\partial_z^2\psi\right|_{x=y=z=0},
\end{equation}
which in particular constrain the form of the Taylor expansion of $\psi$ up to quadratic order. Combining these with similar relations obtained by considering the remaining Lie derivatives, it follows that the Taylor expansion up to quadratic order must depend on $x^2+y^2+z^2=r^2$.

The same analysis for scalar functions involving Lie derivatives can be applied to tensors, including in particular the metric tensor (see, e.g.,~\cite{Rinne:2005df} for a treatment of the axisymmetric case). Alternatively, we can use some other relation to obtain regularity conditions. Consider, for instance, the conservation equation for the fluid~\eqref{eq:cons}: 
\begin{equation}
2(p_r-p_t)+rp'_r+r\left(\rho+p_r\right)\nu'=0.
\end{equation}
Both $p_r$ and $p_t$ must be even parity, so $r p'_r$ is also even parity, and so likewise  $r \nu'$ is even parity, implying $\nu$ itself must be even parity (independent of dimensionality and dynamics).  This makes sense because $\nu(r)$ is in principle directly observable as the gravitational redshift. 

Having thus determined that one of the metric coefficients in Eq.~\eqref{eq:2gstaticchi} must be even parity, it is reasonable to expect that all metric coefficients would satisfy this property. We will not complete a full formal proof of this statement, referring instead to a recent reference focusing on this issue~\cite{Antonelli:2025zxh} (see also~\cite{Zhou:2022yio} for a discussion of the geodesic incompleteness that results from these conditions not being satisfied in specific models). For the Misner--Sharp quasilocal mass defined in Eq.~\eqref{eq:chionshell}, this implies that $m(r)/r^{D-3}$ must be even parity. Hence, the parity of $m(r)$ required for regularity is dimension-dependent.
 
Indeed, consider $\chi = 1- g^{ab} \nabla_a r \nabla_b r = 1 - (1-2m(r)/r^{D-3}) = 2m(r)/r^{D-3}$. 
Since in curvature coordinates $ g^{ab} \nabla_a r \nabla_b r = g^{rr}$, regularity at the origin implies it must be even parity, so $\chi$ must be even parity and $m(r)/r^{D-3}$ must be even parity.

Let us now see what how this relates to the parities of $\bm{\alpha}$ and $\bm{\beta}$, using in particular the ordinary differential equation for $m(r)$, Eq.~\eqref{eq:genmass}. Consider two cases, each with two sub-cases:
\begin{itemize}
\item Inside the star (assuming an analytic core):
\begin{itemize}
\item $D$ even: $\bm{\alpha} + \bm{\beta} \times (odd) = (even)
\quad\Longrightarrow\quad 
\bm{\alpha}=(even) \hbox{ and } \bm{\beta} = (odd) $.

\item $D$ odd: $\bm{\alpha} + \bm{\beta} \times (odd) = (odd)
\quad\Longrightarrow\quad 
\bm{\alpha}=(odd) \hbox{ and } \bm{\beta} = (even)$.
\item 
Overall, $\bm{\alpha}$ has parity $(-1)^D$, 
while $\bm{\beta}$ has parity $(-1)^{D-1}$.
\item 
In either case $\bm{\alpha}$ and $\bm{\beta}$ have opposite parity.
\end{itemize}

\item In the vacuum region outside the star ($\rho\to0$):
\begin{itemize}
\item $D$ even: 
$\bm{\alpha} + \bm{\beta} \times (odd) = (zero)
\quad\Longrightarrow\quad \bm{\alpha} \hbox{ and } \bm{\beta} $ have opposite parity.

\item $D$ odd: 
$\bm{\alpha} + \bm{\beta} \times (odd) = (zero)
\quad\Longrightarrow\quad \bm{\alpha} \hbox{ and } \bm{\beta} $ have opposite parity.
\item In either case $\bm{\alpha}$ and $\bm{\beta}$ have opposite parity, regardless of the dimensionality.
\end{itemize}

\end{itemize}

Note that, while the individual parities of $\bm{\alpha}$ and $\bm{\beta}$ are determined inside matter, in the vacuum region the only statement that can be made at this stage is that $\bm{\alpha}$ and $\bm{\beta}$ must have opposite parity as a necessary condition for geodesic completeness.

Let us consider general relativity in $D=4$ as a consistency check. The Misner--Sharp quasilocal mass (at least inside the star) would then be
\begin{equation}
m(r) = \sum_{n=0}^\infty {\left.m^{(3+2n)}\right|_{r=0}\over(3+2n)!} \; r^{3+2n}.
\end{equation}
This is an odd-parity function of $r$.
But once one goes above the surface of the star $\rho\to0$ and $m(r)\to M$, a constant (even parity). 
On the other hand,
\begin{equation}
\bm{\alpha} = 2 (1-\chi)\to {4 m\over r}; \qquad \bm{\beta} = -2 r.
\end{equation}
So inside the star $m= (odd)$, $\bm{\alpha} = (even)$ and $\bm{\beta}=(odd)$ (which is necessary for geodesic completeness, according to our discussion above), while outside the star \mbox{$m(r)\to M= (even)$}, $\bm{\alpha} = (odd)$ and $\bm{\beta}=(odd)$ which, in view of our discussion above,  
would become incompatible with geodesic completeness if the vacuum region were to extend all they way down to $r=0$.

\section{Solutions of the equations of stellar equilibrium}
\label{Sec:Sols}

In this section we shall solve the equations of stellar equilibrium for specific families of theories in order to illustrate general features of these equations.

\subsection{Overview}

The TOV equation~\eqref{eq:gentov} is valid for any theory of gravity satisfying the master field equations, characterized by the spherically symmetric Einstein tensor being deformed into an identically conserved tensor containing up to second derivatives of the metric~\cite{Carballo-Rubio:2025ntd}. It allows us to describe in a unified formalism a wide range of frameworks discussed before, such as Lovelock gravities~\cite{Kobayashi:2005ch,Maeda:2005ci,Dominguez:2005rt,Ghosh:2008jca,Cai:2008mh} and quasitopological gravities in higher dimensions~\cite{Myers:2010ru,Oliva:2010eb,Bueno:2016xff,Hennigar:2017ego,Ahmed:2017jod,Bueno:2024dgm,Bueno:2024eig,Bueno:2024zsx} (for which the problem of stellar equilibrium has been recently studied~\cite{Bueno:2025tli}), related theories in four dimensions~\cite{Bueno:2025zaj,Borissova:2026wmn,Borissova:2026krh}, as well as effective descriptions of quantum gravity corrections~\cite{Borissova:2026dlz}. In practice, these different theories or frameworks enter through the definition of the functions $\bm{\alpha}(r,\chi)$ and $\bm{\beta}(r,\chi)$.

In this section, our aim is extracting some physical implications of the equations~\eqref{eq:gentov}. Solving these equations explicitly requires making choices along the way, although some properties can be derived invoking general arguments. Hence, we will be treading a fine line between generality and concreteness, using a specific but broad family of theories to illustrate general results.

\subsection{Specific families of theories}

Before starting to break down the study of the equations of stellar equilibrium, we specify the subset of theories we will be using in our exploration.

In particular, all expressions explicitly given below will be particularized for theories satisfying the integrability condition
\begin{equation}\label{eq:constalphabeta}
    \partial_{\chi}\bm{\alpha}-\partial_{r}\bm{\beta}=0,
\end{equation}
in which case there exists a potential function $\bm{\Omega}(r,\chi)$ such that
\begin{equation}\label{eq:alphabetadef}
\bm{\alpha}=\partial_r \bm{\Omega},\qquad\qquad \bm{\beta}=\partial_\chi\bm{\Omega}.
\end{equation}
As discussed above, this family includes general relativity as a particular case. Note also that, for regular geometries, $\bm{\alpha}$ and $\bm{\beta}$ have opposite parity, with the latter having the same parity of $\bm{\Omega}$.

Within this subset of theories, it is useful to define two disjoint families of theories~\cite{Boyanov:2025pes}, which is useful from a mathematical standpoint but is also rooted in historical developments. The first of these two families is the Ziprick--Kunstatter (ZK) family for which, following the definition in~\cite{Boyanov:2025pes}, the potential takes the simple form
\begin{equation}
\bm{\Omega}(r,\chi) = -(1-\chi)\bm{\beta}(r), 
\end{equation}
and, as a consequence,
\begin{equation}
    \bm{\alpha}(r,\chi)=-\left(1-\chi\right)\frac{\text{d}\bm{\beta}(r)}{\text{d}r},
    \qquad \bm{\beta}(r,\chi)=\bm{\beta}(r).
\end{equation}
These theories were introduced, from a different perspective and using a different formalism, by Ziprick and Kunstatter~\cite{Ziprick:2010vb}. These theories can describe well-known regular black holes, in particular the Bardeen regular black hole~\cite{Bardeen1968} (which will be discussed explicitly below) and the Fan--Wang regular black hole~\cite{Fan:2016hvf}, as vacuum solutions.

However the ZK family does not include the well-known Hayward black hole~\cite{Hayward:2005gi}. This was part of the motivation that led Kunstatter, Maeda and Taves to generalize the construction in~\cite{Ziprick:2010vb} for vacuum black holes to a broader class of dilaton theories~\cite{Kunstatter:2015vxa}. The latter construction is equivalent to Horndeski theory (as shown later in, e.g.,~\cite{Takahashi:2018yzc}), and thus the discussion of vacuum solutions therein can be directly incorporated in the formalism used here by constructing a suitable dictionary between both approaches. We define the Kunstatter--Maeda--Taves (KMT) family as having a potential function of the form
\begin{equation}
\bm{\Omega}(r,\chi)=\bm{\Omega}_{\rm GR}(r,\chi)\,\eta\left(\frac{1-\chi}{r^2}\right),   
\end{equation}
with $\eta$ a general function of the indicated argument. The functions $\bm{\alpha}$ and $\bm{\beta}$ can be directly calculated using Eq.~\eqref{eq:alphabetadef}.

The problem of stellar equilibrium in specific instances of the KMT family obtained from the spherically symmetric reduction of polynomial quasitopological gravities has been recently studied~\cite{Bueno:2025tli}. Hence, here we focus on the disjoint ZK family to provide a complementary exploration of the space of theories included in the field equations~\eqref{eq:mfeqs}. Note that both families are treated evenly in the description above in terms of the master field equations. Both families can be lifted to non-polynomial gravitational actions, in particular in the most interesting case of four dimensions~\cite{Borissova:2026wmn,Borissova:2026krh}, which further further supports that neither of these families is preferred over the other based on our current knowledge. It is thus important to study both families either to find aspects which may mark a substantial difference or to corroborate that both families should be treated evenly.

\subsection{Vacuum solutions for ZK theories}

For the ZK class of theories the vacuum equations, Eqs.~(\ref{eq:tteq}-\ref{eq:rreq}), become
\begin{align}
\bm{\alpha}-2\bm{\beta}\frac{\text{d}}{\text{d}r}\left(r^{3-D}m\right)&=0,\label{eq:vacuumcond1st}\\
-\bm{\beta}\nu'-\frac{\bm{\alpha}}{2\left(1-2r^{3-D}m\right)}&=0.\label{eq:vacuumcond2nd}
\end{align}
One way of combining these two equations results in
\begin{equation}
-\bm{\beta}\nu'-\frac{\bm{\beta}}{2\left(1-2r^{3-D}m\right)}\;\frac{\text{d}}{\text{d}r}\left(1-2r^{3-D}m\right)=0,   
\end{equation}
which can be integrated to yield
\begin{equation}
\nu(r)=\nu_0-\frac{1}{2}\ln\left(1-2r^{3-D}m\right),
\end{equation}
where $\nu_{0}$ is a positive constant that can be absorbed in a redefinition of the $t$ coordinate. Hence, the 2-dimensional line element in Eq.~\eqref{eq:2gstatic} becomes
\begin{equation}
q_{ab}(x)\text{d}x^a\text{d}x^b=-\left[1-2r^{3-D}m(r)\right]\text{d}t^2+\frac{\text{d}r^2}{1-2r^{3-D}m(r)},
\end{equation}
where $m(r)$ is determined by Eq.~\eqref{eq:vacuumcond1st}.

Note that all of the resulting metrics satisfy the constraint $g_{tt}\;g_{rr}=-1$. This is a natural $D$-dimensional generalization of a phenomenon that often occurs in (3+1) dimensions~\cite{Jacobson:2007} --- for instance many of the standard regular black holes (Bardeen, Dymnikova, Hayward) are explicitly of this type.
This assumption makes the $tt$ and $rr$ Einstein equations degenerate, and hence the metric only exhibits one degree of freedom. This illustrates the meaning of the constraint $\partial_{\chi}\bm{\alpha}-\partial_{r}\bm{\beta}=0$, and shows that more complicated metrics with $g_{tt}\;g_{rr}\neq -1$ are not compatible with this constraint without the inclusion of matter. This is the reason why, for simplicity, we will restrict ourselves to this subset of theories and geometries, although the calculations below can be generalized to situations in which $\partial_{\chi}\bm{\alpha}-\partial_{r}\bm{\beta}\neq0$.

Another way of combining  Eqs.~(\ref{eq:tteq}-\ref{eq:rreq}) for ZK theories in vacuum is to note
\begin{equation}
-\frac{\text{d}}{\text{d}r}\left(2\bm{\beta} r^{3-D}m\right) = 0.
\end{equation}
This integrates to
\begin{equation}\label{Eq:beta_ZKB}
{\bm{\beta}} = {K r^{D-3}\over 2 m}. 
\end{equation}
This implies that any spacetime metric of the general form 
\begin{equation}
ds^2 =-\left[1-2r^{3-D}m(r)\right]\text{d}t^2+\frac{\text{d}r^2}{1-2r^{3-D}m(r)} + r^2 d\Omega^2
\end{equation}
can be interpreted as the vacuum solution of some specific nontrivial ZK model with dynamics governed by
\begin{equation}
{\bm{\beta}}(r) = {K r^{D-3}\over 2 m(r)}; 
\qquad  {\bm{\alpha}}(r,\chi)= (1-\chi) \partial_r {\bm{\beta}}; 
\qquad {\bm{\Omega}}(r,\chi)= - (1-\chi) {\bm{\beta}}(r).
\end{equation}
Note that the function $\bm{\beta}(r)$ above has even parity by construction.

These expressions allow us to make explicit how the ZK family of theories includes solutions describing regular black holes and, in particular, a suitable $D$-dimensional generalization of the Bardeen black hole, for the choice:
\begin{equation}
\bm{\beta}_{\rm B}^{(D)}(r)=-\frac{D-2}{r^{2(D-3)}}\left(r^2+\ell^2\right)^{3(D-3)/2},
\end{equation}
which, integrating Eq.~\eqref{eq:vacuumcond1st}, results in the expression
\begin{equation}
    f^{(D)}_{\rm B}(r)=1-\frac{4M r^{2\left(D-3\right)}}{\left(D-2\right)\left(r^2+\ell^2\right)^{\frac{3}{2}\left(D-3\right)}}.
\end{equation}
It is straightforward to show that this reduces, for $D=4$, to the well-known Bardeen spacetime~\cite{Bardeen1968}, while for $\ell\to 0$ it reduces to the Tangherlini ($D$-dimensional Schwarzschild) spacetime. Other $D$-dimensional generalizations of the Bardeen space are possible (see for instance~\cite{Kumar:2018vsm}) and we are choosing the one above following the parity criteria that, as described in Sec.~\ref{Sec:SteEqu}, guarantee geodesic completeness.

Note the somewhat subtle nature of the construction: We have explicitly built a particular ZK dynamical model whose ``vacuum solution'' is the Bardeen regular black hole --- a spacetime that is certainly not a ``vacuum solution'' if viewed through the lens of standard general relativity. This process can be interpreted as an example of ``reverse engineering'' the dynamics to generate, as a ``vacuum'' spacetime, a model originally developed for quite different reasons~\cite{Ziprick:2010vb,Kunstatter:2015vxa,Carballo-Rubio:2025ntd}.

\subsection{Curvature regularity conditions}

Not every choice of $\bm{\beta}$ in the ZK family will be compatible with a regular spacetime. 
The only non-zero components of the $D$-dimensional Riemann tensor are
\begin{align}
R_{abcd}&=\mathcal{R}_{abcd},\nonumber\\
R_{aibj}&=-r\gamma_{ij}\nabla_a\nabla_b r\nonumber\\
R_{ijkl}&=r^2\left[1-\left(\nabla r\right)^2\right]\left(\gamma_{ik}\gamma_{jl}-\gamma_{il}\gamma_{jk}\right),
\end{align}
where $R_{abcd}$ is the 2-dimensional curvature tensor.\footnote{We have made use of the expressions in the Appendix of~\cite{Maeda:2007uu}, adapted to our notation.} The $D$-dimensional Kretschmann scalar $\mathcal{K}$ can be written as 
\begin{equation}
\mathcal{K}=2\mathcal{R}^2
+\frac{D-2}{r^2}\;\; \nabla_a\nabla_b r\;\; \nabla^a\nabla^br
+\frac{2(D-2)(D-3)}{r^4}\left[1-\left(\nabla r\right)^2\right]^2,  
\end{equation}
We take the line element~\eqref{eq:2gstatic} and the mass function~\eqref{Eq:beta_ZKB}, i.e.,
\begin{eqnarray}
    m=-\frac{(D-2) r^{D-3}}{\bm{\beta}}m_0,
\end{eqnarray}
where $K=-(D-2)m_{0}/2$ so the Misner-Sharp mass reduces to the ADM mass at infinity. The Kretschmann scalar becomes
\begin{align}
\mathcal{K}=&
\frac{8\left(D-2\right)^{3}\left(D-3\right)m_{0}^2}{r^4\bm{\beta}^2}+\frac{4\left(D-2\right)^{3}\bm{\beta}'^{2}m_{0}^2}{r^2\bm{\beta}^4}+\frac{4\left(D-2\right)\left[2\left(D-2\right)m_{0}+\bm{\beta}\right]^{2}\nu'^2}{r^2\bm{\beta}^2}+\nonumber\\
&
+4\left\{\left[2\left(D-2\right)m_{0}+\bm{\beta}\right]\left(\nu'^{2}+\nu''\right)\bm{\beta}-\left(D-2\right)m_{0}\nu'^{2}\bm{\beta}'\right\}^{2}\bm{\beta}^{-4}.
\label{E:kretchmann}
\end{align}

From the Kretschmann scalar, we can deduce, by looking at the terms that depend separately on $m$ and $\nu$, that, provided $m_0\neq 0$, the conditions required for regularity are as follows: Assuming the following $r\to0$ expansions for $\beta$ and $\nu$
\begin{equation}\label{eq:betamuexps}
    \bm{\beta}(r)=\sum_{m}^{\infty}b_{i}r^{i},\quad \nu(r)=\sum_{0}^{\infty}\mu_{j}r^{j},
\end{equation} 
and inserting  them in~\eqref{E:kretchmann} we find $r=0$ is regular (for vacuum ZK theories) if $m\leq-2$ and $\mu_{1}=0$. 
Note that $m\geq-1$ and $m\geq0$ correspond to spacetimes with so-called integrable singularities~\cite{Casadio:2023iqt,Arrechea:2025fkk}, in which the Misner-Sharp mass vanishes at $r=0$ but the Kretschmann scalar still diverges.

As it was discussed in detail in Section~\ref{Sec:SphSym}, to guarantee geodesic completeness on top of the finiteness of curvature invariants, the spacetime metric needs to be symmetric under $|r|\to -|r|$ transformations. In vacuum, the $\bm{\beta}$ function needs to obey an even-parity expansion near $r=0$, while in presence of matter it needs to be odd. A sensible theory of gravity must be compatible with dilute fluid spheres in equilibrium \textit{before} imposing any regularity condition on its vacuum solutions. Hence, the rest of our discussion will be focused on matter configurations and the odd-parity $\bm{\beta}$ functions compatible with them. The exterior metrics to these stars will not generally correspond to known regular black hole spacetimes, but will share most of their properties.

\subsection{Mass function for ZK theories and constant density}

Let us now turn our attention to the simplest model with matter, 
namely isotropic stars with constant density, the generalization of Schwarzschild's 4-dimensional general-relativistic constant density star. 
In the presence of matter, we can write Eq.~\eqref{eq:genmass} as
\begin{equation}
-\frac{\text{d}}{\text{d}r}\left(2\bm{\beta} r^{3-D}m\right) =16\pi r^{D-2}\rho.
\end{equation}

For constant-density stars, $\rho\to\rho_0$, this equation can be integrated explicitly to give
\begin{equation}\label{Eq:MassZKCD}
m=-\frac{(D-2) r^{D-3}}{\bm{\beta}}m_0-\frac{8\pi r^{2(D-2)}}{(D-1)\bm{\beta}}\rho_0,    
\end{equation}
where the integration constant $m_0$ has been again chosen to reduce to the ADM mass for $\rho_0=0$ and $\bm{\beta}\to \bm{\beta}_{GR}$, but is otherwise a free parameter that might in principle take values other than $M$.
Notice that the term proportional to $\rho_{0}$ always decays near $r=0$ faster than the term proportional to $m_{0}$, so the $m_0$ term, if present, will dominate the contribution to the Misner--Sharp mass at small distances.

For completeness, we show the $D$-dimensional TOV equation for the ZK family of functions
\begin{equation}\label{eq:TOVZK}
    p'=\frac{\rho_{0}+p}{r \bm{\beta}}\left[\frac{8\pi \bm{\beta} r^D p  -m_{0} r^2\bm{\beta}'+8\pi \bm{\beta}' r^{D+1}\rho_{0}/\left(D-1\right)}{r\bm{\beta}-2m_{0}+16\pi r^D\rho_{0}/\left(D-1\right)}\right]
\end{equation}
which cannot be analytically solved for the cases that we will consider below. 
Note that this is the isotropic TOV appropriate for a perfect fluid introduced into a dynamical ZK model with non-trivial ``vacuum'' solutions. 

From here onward, we will restrict our discussion to $D=4$. In most situations, we will want solutions to the TOV equation to have a smooth limit to general relativity. Hence we will impose $m_{0}=0$ in the region containing the perfect fluid. The resulting TOV equation admits the simple form
\begin{eqnarray}\label{Eq:TOVbeta}
    p'=\frac{\rho_{0}+p}{r\bm{\beta}}\left(\frac{8\pi r^{3}p-m\bm{\beta}'}{1-2m/r}\right).
\end{eqnarray}

We have shown previously our general discussion on parity requirements in Sec.~\ref{sec:Parity} that the geodesic completeness of stars at $r=0$ requires $\bm{\beta}$ to have odd parity. This is made explicit for the ZK family by Eq.~\eqref{Eq:MassZKCD} (with $m_0=0$). Hence, we will focusing on specific examples satisfying this parity requirement below.

\subsection{A family of ZK theories with regular four-dimensional matter configurations}

\begin{figure}
    \centering
    \includegraphics[width=0.6\linewidth]{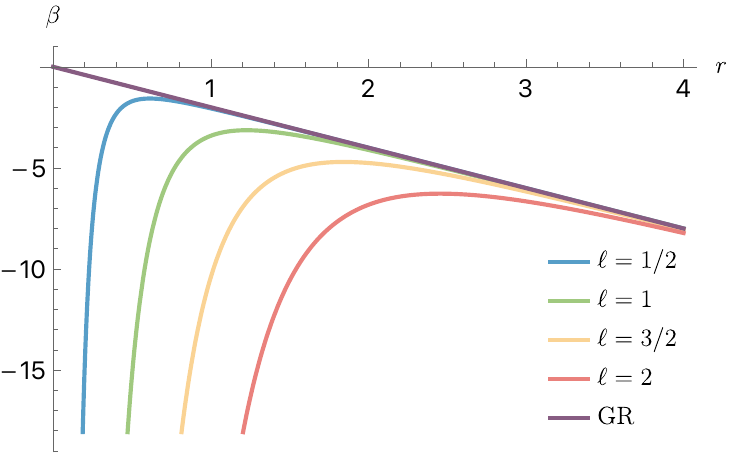}
    \caption{Plot of the $\bm{\beta}_{\text{odd}}$ function with $n=1$ in terms of $r$. This function is constructed to always have a global maximum for $\ell>0$. As $\ell\to0$, this function approaches its general relativity behavior from below. As $n$ is increased, the $r\to0$ divergence becomes stronger, but the $r/\ell\gg1$ behaviour is unaffected.}
    \label{Fig:Beta}
\end{figure}

For our analysis of stars, we restrict ourselves only to $\bm{\beta}$ functions that obey odd expansions at the origin in four dimensions. This choice guarantees that the Misner-Sharp mass has the required odd-parity behavior in matter at the expense of making vacuum solutions incomplete. Since we have complete freedom to select such a family of functions, we propose the following illustrative example
\begin{eqnarray}\label{eq:Betaodd}
    {\bm{\beta}}_{\rm odd}=-2r-\frac{2\ell^2}{\sqrt{r^2+\ell^2}}\left(\frac{\ell}{r}\right)^{2n+1},
\end{eqnarray}
where the index $n$ controls the power of the divergence in $\bm{\beta}_{\text{odd}}$ at $r=0$. This function has the appealing properties of reducing to the appropriate form in general relativity ($\bm{\beta}_{\rm GR}=-2r$) in the $\ell\to0$ and $r/\ell\gg1$ limits, diverging at the origin as \mbox{$\bm{\beta}_{\text{odd}}\propto-2\ell^{2n+2}/r^{2n+1}$}, and being bounded from above by $\bm{\beta}_{\rm GR}$. These are all 
rather generic properties shared by all $\bm{\beta}$ functions that regularize the vacuum solutions of general relativity. 
Figure~\ref{Fig:Beta} shows a plot of the $\bm{\beta}_{\text{odd}}$ function for $n=1$ and different values of $\ell$.

The Misner-Sharp mass associated to this choice (assuming a constant-density fluid) is
\begin{eqnarray}
m_{\text{odd}}=\frac{m_{0}r\sqrt{r^2+\ell^2}}{r\sqrt{r^2+\ell^2}+\ell^{2}\left(\ell/r\right)^{2n+1}}+\frac{4\pi \rho_{0}r^4\sqrt{r^2+\ell^2}}{3\left[r\sqrt{r^2+\ell^2}+\ell^2\left(\ell/r\right)^{2n+1}\right]}.
\end{eqnarray}
For $n=0$, we get a Misner-Sharp mass that contains an integrable singularity in vacuum ($\rho_{0}=0$), while for $n$ greater than one and $m_{0}=0$, the Misner-Sharp mass contains only odd powers of $r$ and has a finite Kretschmann invariant. Increasing $n$ causes a faster vanishing of the Misner-Sharp mass as $r=0$ is approached, which, in turn, will influence the behavior of the pressure through the TOV equation~\eqref{Eq:TOVbeta}. In the next subsections, we will analyze some generic properties of the fluid pressure in these models.

\subsection{Integrating the pressure: regular solutions}
Let us analyze how the contribution form $\bm{\beta}$ will affect hydrostatic equilibrium configurations and their compactness limits. We stress that the bounds obtained below are generic and satisfied by any $\bm{\beta}$ function that regularizes curvature singularities in vacuum. For the choice~\eqref{eq:Betaodd}, we have 
\begin{equation}\label{eq:betalessGR}
\bm{\beta}_{\text{odd}}\leq \bm{\beta}_{\rm GR},
\end{equation}
and
\begin{equation}
|\bm{\beta}_{\text{odd}}|\geq |\bm{\beta}_{\rm GR}|.    
\end{equation}
Therefore, it follows that
\begin{equation}
m_{\text{odd}}\leq m_{\rm GR} =\frac{4\pi r^{3}}{3}\rho_0.
\end{equation}
A star with constant density $\rho_{0}$ and same radius $R$ has therefore less mass when $\bm{\beta}=\bm{\beta}_{\text{odd}}$ than in the general relativity case. Hence,
\begin{equation}
1-2r^{-1}m_{\text{odd}}(r)\geq 1-r^{-1}m_{\rm GR}=  1-\frac{8\pi r^2}{3}\rho_0.   
\end{equation}
On the other hand, we have
\begin{eqnarray}
    \bm{\beta}_{\text{odd}}'>\bm{\beta}_{\rm GR}'
\end{eqnarray}
Putting all of this together in the TOV equation~\eqref{Eq:TOVbeta}, we have
\begin{eqnarray}\label{Eq:TOVbeta2}
    p'=\frac{\rho_{0}+p}{r\bm{\beta}_{\text{odd}}}\left(\frac{8\pi r^{3}p-m_{\text{odd}}\bm{\beta}_{\text{odd}}'}{1-2m_{\text{odd}}/r}\right)\geq \frac{\rho_{0}+p}{r\bm{\beta}_{\rm GR}}\left(\frac{8\pi r^{3}p-m_{\rm GR}\bm{\beta}_{\rm GR}'}{1-2m_{\rm GR}/r}\right).
\end{eqnarray}
We see that, at the star's surface $r=R$ (where $p(R)=0$), pressure initially tends to grow inwards more slowly than in general relativity (i.e., with a less negative gradient). As the TOV equation is integrated to smaller radii, its departure from the solution in general relativity grows due to the change in sign of the $m\bm{\beta}'$ term when $r/\ell<1$.
Summarizing, we have
\begin{equation}
    \frac{\text{d}p_{\text{odd}}}{\text{d}r}\bigg|_{r=R}\geq \frac{\text{d}p_{\rm GR}}{\text{d}r}\bigg|_{r=R}
\end{equation}
and
\begin{equation}
    p_{\text{odd}}\leq p_{\rm GR}.
\end{equation}
We can understand this result as an indication that, at least for static situations, $\bm{\beta}$ dictates the strength of the gravitational interaction. For the theories we are considering, the constraint Eq.~\eqref{eq:betalessGR} indicates that this strength is weaker everywhere (thus resulting in a weaker pressure profile).

Now let us show this explicitly by numerically integrating the TOV equation~\eqref{Eq:TOVbeta} which, for $n=1$, results in 
\begin{equation}\label{eq:TOVn3}
    p'=
    \frac{4\pi r\left(\rho_{0}+p\right)\sqrt{r^2+\ell^2}}{r\sqrt{r^2+\ell^2}\left(8\pi r^2 \rho_{0}-3\right)-3\ell^{5}/r^4}\left\{3p+\rho_{0}\left[1-\frac{\ell^5\left(5r^2+4\ell^2\right)}{\left(r^2+\ell^2\right)\left(r^4\sqrt{r^2+\ell^2}+\ell^5\right)}\right]\right\},
\end{equation}
where one can clearly see that the $\bm{\beta}'_{\text{odd}}$ term contributes negatively to the numerator. This equation can be checked to have the correct general relativity limit for $\ell\to0$, i.e.,
\begin{equation}\label{Eq:TOVl0}
    p' =\frac{4\pi r\left(\rho_{0}+p\right)\left(3p +\rho_{0}\right)}{\left(8\pi  r^2\rho_{0}-3\right)}+\mathcal{O}\left(\ell^2\right).
\end{equation}

We want to integrate~\eqref{eq:TOVn3} from a surface of vanishing pressure at $r=R$ inwards. The Misner-Sharp mass takes the form
\begin{align}
    m(r)=
    &
    M\left[\frac{r^4\sqrt{r^2+\ell^2}}{r^4\sqrt{r^2+\ell^2}+\ell^5}\right],
    \quad~~ r>R,\nonumber\\
    m(r)=
    &
    \frac{4}{3}\pi \rho_{0}r^3 \left[\frac{r^4\sqrt{r^2+\ell^2}}{r^4\sqrt{r^2+\ell^2}+\ell^5}\right], 
    \quad r\leq R.
\end{align}

The junction conditions for the field equations~\eqref{eq:feqs} result in the following relations (see App.~\ref{app:junc}):
\begin{eqnarray}\label{Eq:JunctionR}
    \lim _{r\to R^{+}}m(r)-\lim _{r\to R^{-}}m(r)=0,\quad \lim _{r\to R^{+}}\nu'(r)-\lim _{r\to R^{-}}\nu'(r)=0.
\end{eqnarray}

Continuity of the mass function at $r=R$ requires 
\begin{equation}\label{eq:critrho}
    \rho_{0} = \frac{3M}{4\pi R^3}.
\end{equation} 
The conservation relation~\eqref{eq:cons} can be directly integrated for the constant-density star
\begin{equation}\label{eq:psolcd}
    p(r)=-\rho+e^{-\nu(r)+\mu_{0}}\left(p_{0}+\rho\right),
\end{equation}
where $p_{0}$ is an integration constant that selects the pressure at $r=0$.

\begin{figure*}[htp] 
    \centering
    \subfloat{%
        \includegraphics[width=0.5\linewidth]{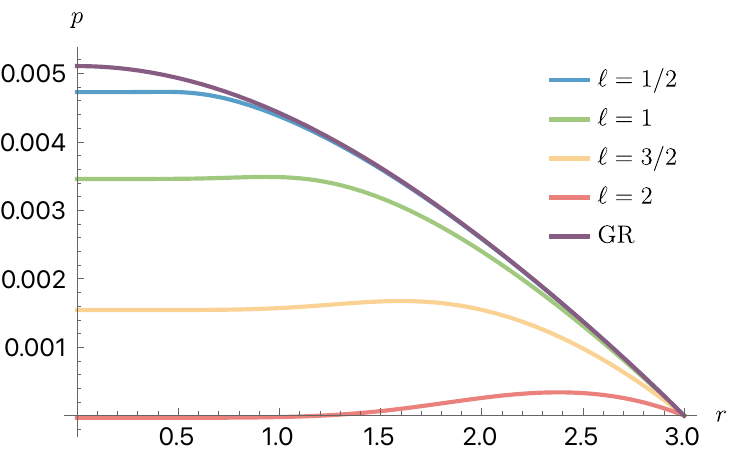}%
        \label{fig:a1}%
        }%
    \hfill%
    \subfloat{%
        \includegraphics[width=0.5\linewidth]{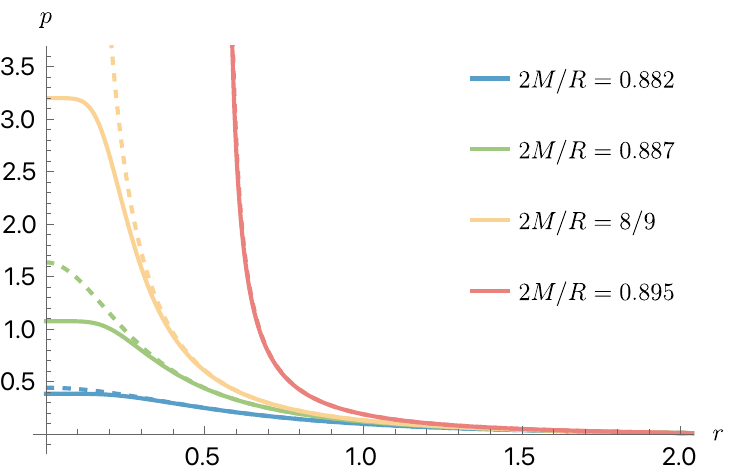}%
        \label{fig:b1}%
        }%
    \caption{Left panel: Pressure profile of constant-density stars with $M=1$ and $R=3$. As $\ell$ is increased, the central pressure decreases. All solutions lie below the general relativity one. Right panel: Pressure profiles of constant-density stars with varying compactness and $M=1$. The dashed lines correspond to the general relativity solutions, and the continuous ones to $\ell=1/5$. The general relativity Buchdahl limit corresponds to $2M/R=8/9$. At the Buchdahl compactness, the general relativity pressure diverges at $r=0$, while $\ell=1/5$ solution is regular.}
\label{Fig:PressureWithL}
\end{figure*}

We can assume expansions for $\bm{\beta}$ and $p$ that are consistent with geodesic completeness at $r=0$ (i.e., strictly odd and even functions, respectively),
\begin{equation}\label{eq:betapexp}
\bm{\beta}(r)=\sum_{m}^{\infty}b_{2i+1}r^{2i+1},\quad p=\sum_{j=0}^{\infty}p_{j}r^{j},
\end{equation}
Replacing them in the TOV equation, we obtain, expanding for small $r$,
\begin{align}\label{eq:tovexp}
&
2p_{2}r+\mathcal{O}\left(r^3\right)=
-\frac{8\pi r^2\left(\rho_{0}+p_{0}\right)\left[
 \left(2m+1\right)\rho_{0}+3p_{0}\right]+\mathcal{O}\left(r^3\right)}{3\left[b_{2m+1}+b_{2m+3}r^{2}+\mathcal{O}(r^4)\right]r^{2m+1}+\mathcal{O}\left(r^{3}\right)},
\end{align}
where $m$ is now a \textit{negative} integer denoting the power of the strongest divergent term in \mbox{$\bm{\beta}\propto b_{2m+1}r^{2m+1}$}. We see that, for $m\leq -1$ the denominator has a divergent contribution proportional to $r^{2m+1}$.
The stronger the divergence in $\bm{\beta}$, the faster the pressure will approach a constant value $p_{0}$ as $r\to0$.
Particularizing for the $\bm{\beta}_{\text{odd}}$ function with $n=1$ (equivalent to taking $m=-2$ and $b_{-3}=-2\ell ^4$ in~\eqref{eq:betapexp}), we have
\begin{align}\label{eq:tovexp2}
2p_{2}r+4p_{4}r^3+6p_{6}r^5+\mathcal{O}(r^6)=
\frac{4\pi r^5}{\ell^4} \left(\rho_{0}+p_{0}\right)\left(\rho_{0}-p_{0}\right)+\mathcal{O}\left(r^{7}\right).
\end{align}
We see that, in this case, the TOV equation further enforces $p_{2}=p_{4}=0$, and the first non-zero coefficient is
\begin{eqnarray}
    p_{6}=\frac{2\pi }{3 \ell^4}\left(\rho_{0}^2-p_{0}^2\right).
\end{eqnarray}
The remaining $p_{2j}$ coefficients are straightforward to obtain order-by-order. This is a direct consequence of the divergence in $\bm{\beta}$~\eqref{eq:Betaodd}, which tends to ``flatten" the pressure near $r=0$. The left panel in Figure~\ref{Fig:PressureWithL} shows some example numerical integrations showing how increasing $\ell$ decreases the central pressure, even to negative values.

\subsection{Behavior of solutions with divergent pressure}

This taming effect on the growth of the pressure does not prevent its eventual divergence if the star is compact enough.
Solutions with divergent pressures at $r=0$ still exist. For constant density stars in GR, these solutions act as the separatrix between regular and singular stars, indicating the presence of a maximum compactness limit~\cite{PhysRev.55.413,PhysRev.55.364,Arrechea:2021pvg}.We can find the leading-order contributions to the pressure assuming now expansions of the form
\begin{equation}
  \bm{\beta}(r)=\sum_{m}^{\infty}b_{2i+1}r^{2i+1},\quad p(r) = \sum_{k}^{\infty}p_{j}r^{j}.
\end{equation}
Replacing them in the TOV equation we find the leading-order divergence in the pressure to be
\begin{equation}\label{eq:Psep}
    p_{k}=\frac{\left(m-1\right)}{4\pi}b_{2m+1},\quad k=2\left(m-1\right).
\end{equation}
The general relativity result, $p_{2}=1/2\pi r^2$, is recovered by setting $m=0$ and $b_{1}=-2$. For $m=-2$ and $b_{-3}=-2\ell^4$ (corresponding to $\bm{\beta}_{\text{odd}}$ with $n=1$), hence the leading term is
\begin{equation}
    p_{-6}=\frac{3\ell^4}{2\pi},\quad k = -6.
\end{equation}
This is the dominant behavior of the pressure near $r=0$ when the compactness of the star is equal to the $\ell$-dependent Buchdahl limit. Beyond that value, the divergence in the pressure appears at some $r>0$ surface that will move outwards as the star is compressed. Assuming that $\bm{\beta}$ takes a constant value at some $r=r_{\text{Div}}$, we find a solution of the form
\begin{equation}
   \bm{\beta}(r)=\sum_{i=0}^{\infty}b_{i}\left(r-r_{\text{Div}}\right)^{i},\quad  p(r)=\sum_{j=-1}^{\infty}q_{j}\left(r-r_{\text{Div}}\right)^{j},
\end{equation}
with 
\begin{equation}
    q_{-1}=-\frac{2r_{\text{Div}}\rho_{0}}{3}-\frac{b_{0}}{8\pi r_{\text{Div}}^2},
\end{equation}
which is positive as long as $b_{0}<-16\pi r_{\text{Div}}^3\rho_{0}/3$.
For $\bm{\beta}_{\text{odd}}$ case, we obtain
\begin{equation}
    q_{-1}=-\frac{2r_{\text{Div}}\rho_{0}}{3}+\frac{1}{4\pi r_{\text{Div}}}+\frac{1}{4\pi \sqrt{r_{\text{Div}}^2+\ell^2}}\left(\frac{\ell}{r}\right)^{3+2n},
\end{equation}
which reduces to the general relativity result in the $\ell\to0$ limit. 

The divergence in the pressure as $p\propto \left(r-r_{\text{Div}}\right)^{-1}$ is almost independent of the behaviour of $\bm{\beta}$. This divergence is ruled by the large-pressure regime of the TOV equation~\eqref{eq:TOVZK} at some $r_{\text{Div}}>0$, where $\bm{\beta}$ just corrects the general relativity solution perturbatively. The divergence at $r=0$ obtained in~\eqref{eq:Psep}, however, is very sensitive to $\bm{\beta}$ since this function dramatically affects the short-$r$ behaviour of the equations of motion. The right panel in Figure~\ref{Fig:PressureWithL} shows how divergence is reached for a family of stars with $\ell=1/5$. The corrected solution always lies below its general relativity counterpart (in dashed lines). The yellow curve shows how there are regular solutions that exceed the Buchdhal compactness once corrections due to $\ell$ are present.

\subsection{Regular black holes with fluid cores}

By the above analysis we conclude that, quite generically, gravity theories whose vacuum solutions are regularized by the introduction of some length-scale will lead to a relaxation of the bounds imposed by general relativity on the compactness of fluid spheres, in accordance with the complementary analysis in~\cite{Bueno:2025tli}. Another generic consequence of regularizing the vacuum solutions of the theory is the introduction of a second, inner horizon whose radial position is controlled by the scale $\ell$. Beyond this inner horizon lies a region where it is possible to place fluid spheres in equilibrium just as we can do in the region outside the outer horizon. This phenomenon is already present in general relativity for situations in which a charge $Q$ is present~\cite{PoncedeLeon:2017usu}, as the latter induces the existence of an inner horizon as well~\cite{Penrose:1968ar,Matzner:1979zz,Poisson:1989zz,Poisson:1990eh,Ori:1991zz}. These solutions were described for the KMT family in~\cite{Bueno:2025tli}, and that these exist also for the ZK family is consistent with the existence of the inner horizon being the critical necessary ingredient. As described below, the features of the solutions with perfect fluid cores of theories with $\ell\neq0$ are similar to the features solutions with $Q\neq 0$ in general relativity.

Since vacuum solutions belonging to the family specified by $\bm{\beta}_{\rm odd}$ are not $|r|\to-|r|$ symmetric at $r=0$, only geometries where the centre $r=0$ is covered by a perfect fluid are consistent, even if said fluid lies deep inside the inner horizon and is invisible to outside observers. One might interpret these configurations as possible end-states of gravitational collapse. However, the known instability of inner horizons, which would be Cauchy horizons in this case, must be contemplated in this picture~\cite{Penrose:1968ar,Matzner:1979zz,Poisson:1989zz,Poisson:1990eh,Ori:1991zz,Carballo-Rubio:2018pmi,Carballo-Rubio:2019nel,Carballo-Rubio:2021bpr,Carballo-Rubio:2022kad,Carballo-Rubio:2024dca}. 

\begin{figure}
    \centering
    \includegraphics[width=0.65\linewidth]{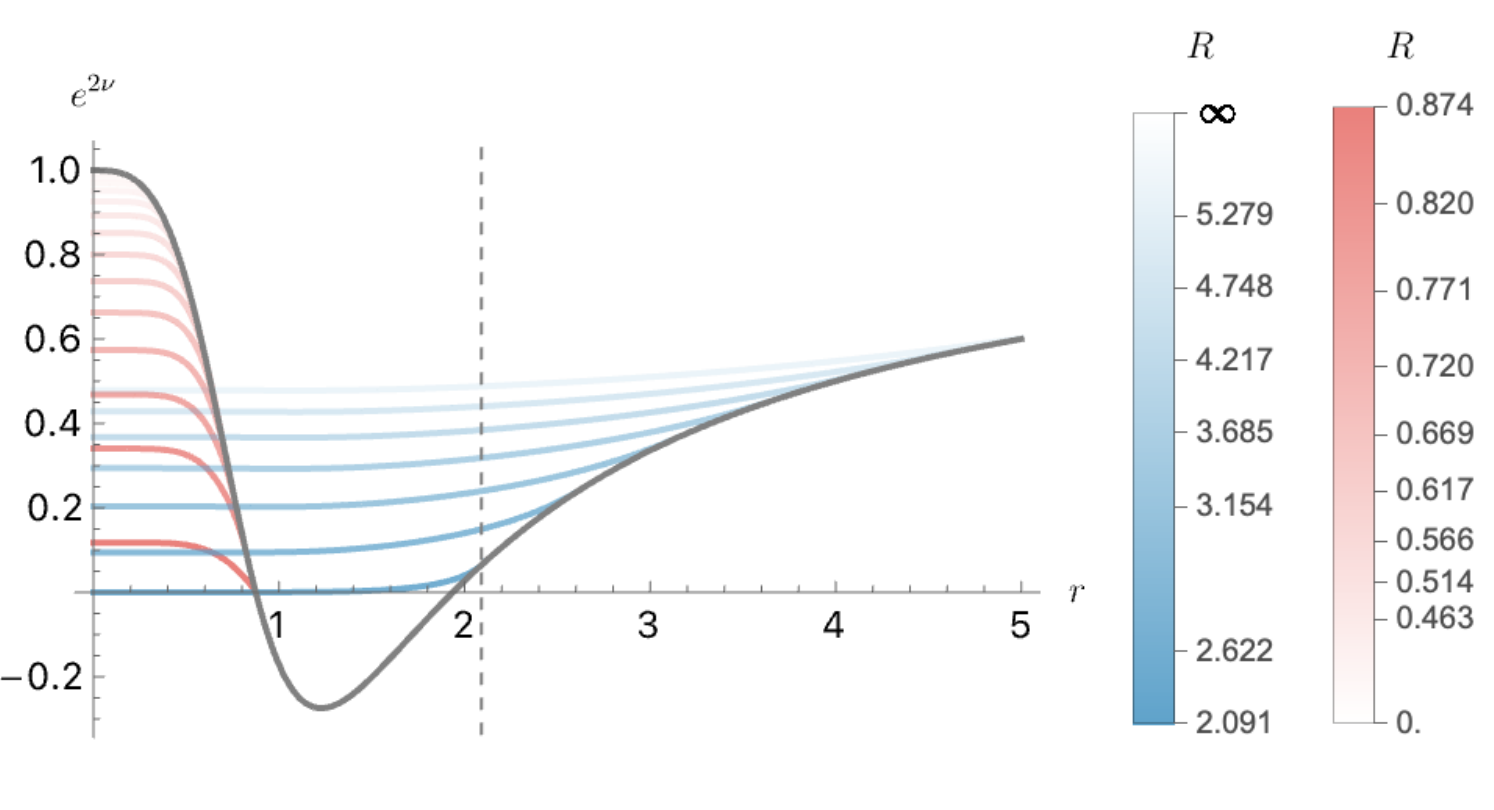}
    \caption{Redshift functions of families of constant-density stars of varying compactness for a fixed value of $\ell$ and $M=1$. The dark gray curve is the redshift function from the vacuum solution with $\bm{\beta}=\bm{\beta}_{\text{odd}}$ with $n=1$ and $\ell=1$, and the region in between horizons shaded in light gray. 
    The surface of constant-density stars, defined by the condition $p(R)=0$, can be placed outside the outer horizon (in blue), or inside the inner horizon (in red). Stars with their surface outside the inner horizon have an inwards-decreasing redshift that, in the limiting case corresponding to the Buchdahl limit, vanishes at $r=0$ (darkest blue curve) for $2m(R)/R\approx0.935$ (its surface is indicated by the dashed, vertical line). Stars with their surface inside the inner horizon have an inwards-increasing redshift that is always finite for any compactness between $0$ and $1$. In our units, the radial positions of the outer and inner horizons are $r_{\text{out}}\approx1.936$ and $r_{\text{in}}\approx0.874$, respectively.}
    \label{Fig:SmallStars}
\end{figure}

\begin{figure*}[htp] 
    \centering
    \subfloat{%
        \includegraphics[width=0.5\linewidth]{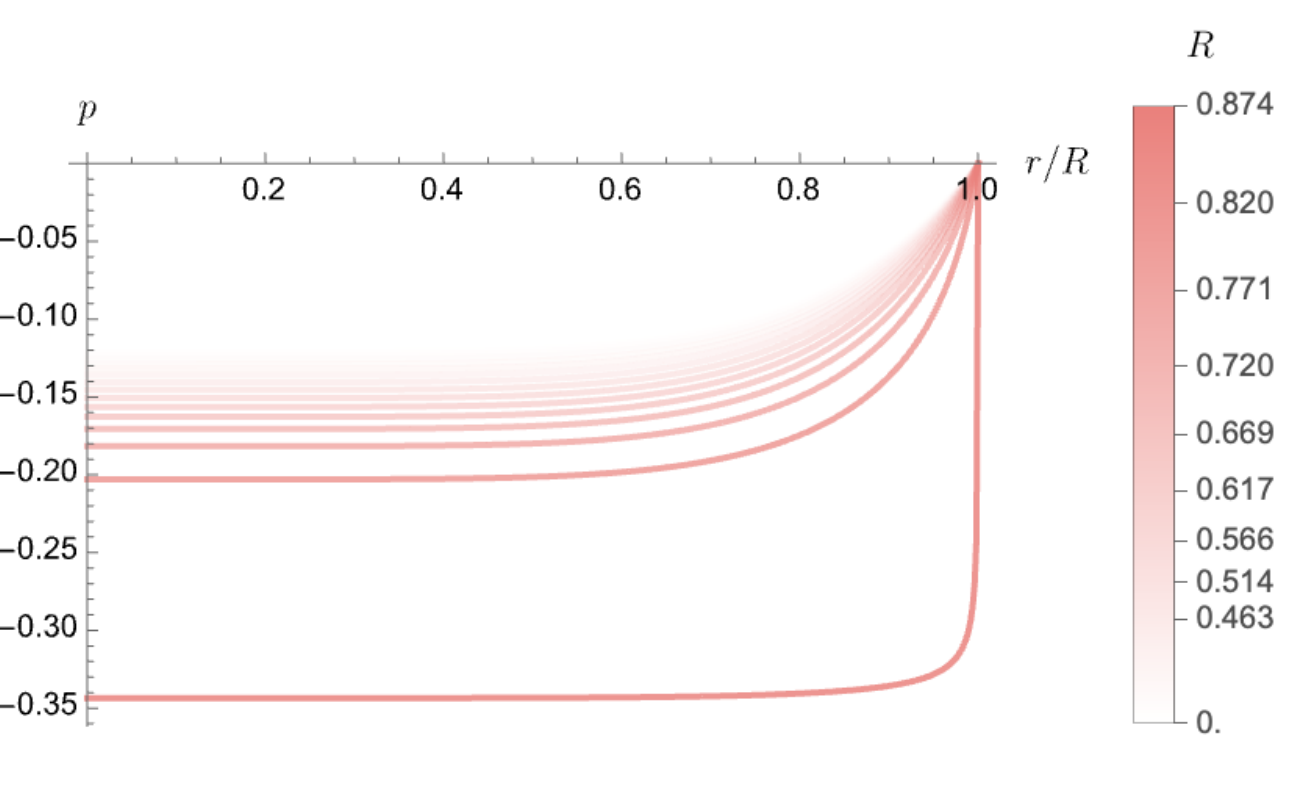}%
        \label{fig:a3}%
        }%
    \hfill%
    \subfloat{%
        \includegraphics[width=0.5\linewidth]{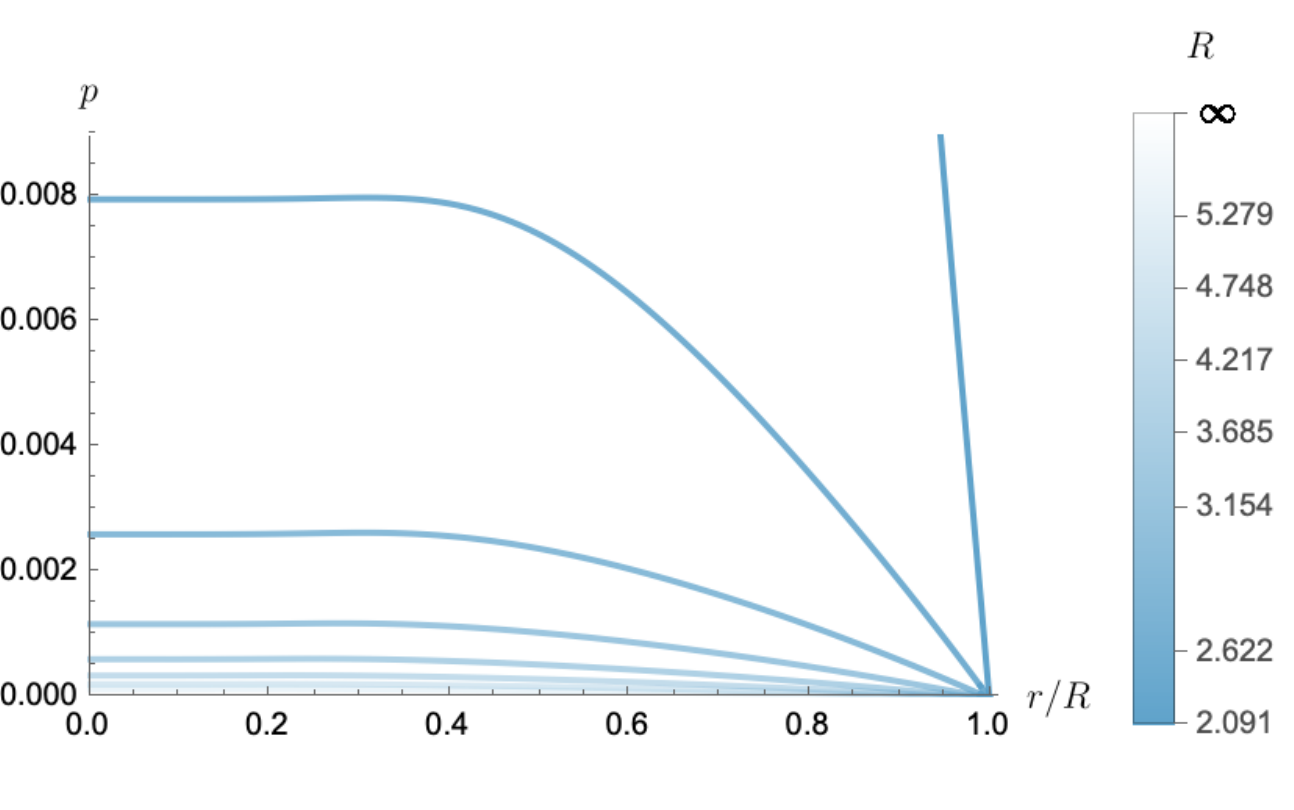}%
        \label{fig:b3}%
        }%
    \caption{Further details concerning the solutions shown in Fig.~\ref{Fig:SmallStars}.
    \textit{Left panel:}
    Pressure profiles for stars with their surfaces placed inside the inner horizon of the vacuum solution. Stars attached closer to the inner horizon (from inside) have pressures that decrease faster towards the interior, approaching $p=-\rho_{0}$ in the limit in which the surface becomes coincident with the inner horizon. \textit{Right panel:} Pressure profiles for stars with their surfaces placed outside the outer horizon. Stars attached closer to the outer horizon (from outside) have pressures that become divergent in the limiting solution with compactness given by the Buchdahl-like limit, which in the theories we are working depends on $\ell$.}
    \label{Fig:PressuresSmall}
\end{figure*}

Figures~\ref{Fig:SmallStars} and~\ref{Fig:PressuresSmall} display the redshift function and pressure profiles of constant-density stars placed both outside the outer horizon (in blue) and inside the inner horizon (in red) for a $\bm{\beta}_{\text{odd}}$ function with $\ell=1$ and $n=1$. Darker blue (red) lines correspond to stars closer to the outer (inner) horizons. In Fig.~\ref{Fig:SmallStars}, The thick gray curve is the vacuum redshift function, which needs to be matched with one of the colored lines in order to construct a well-defined spacetime. The region in-between the horizons, where no static fluid sphere can be placed, has been shaded in gray.

Let us start by considering fluid spheres placed outside the outer horizon. For convenience, we recall here the TOV equation
\begin{eqnarray}\label{Eq:TOVbeta3}
    p'=\frac{\rho_{0}+p}{r\bm{\beta}}\left(\frac{8\pi r^{3}p-m\bm{\beta}'}{1-2m/r}\right).
\end{eqnarray}
Just below the surface, these stars have inward-increasing pressures. We can sketch the behaviour of the pressure just from the various terms appearing in~\eqref{Eq:TOVbeta3}, and by realizing that
\begin{eqnarray}\label{Eq:TOV2}
    \frac{d}{dr}\left(e^{2\nu(r)}\right)=\frac{d}{dr}\left(1+\frac{4m_{0}}{\bm{\beta}}\right)=-\frac{4m_{0}\bm{\beta}'}{\bm{\beta}^2},
\end{eqnarray}
that is, the sign of $\bm{\beta}'$ is minus the sign of $\nu'$, which further implies that, at the surface $p(R)=0$ of stars outside the outer horizon (where $\nu'>0$, i.e., $\bm{\beta}'<0$) the numerator of the TOV equation is always negative due to the $-m\bm{\beta'}$ term, and pressure will always grow inwards. However, it will do so with less slope than in general relativity, due to the fact that $\bm{\beta}'\geq\bm{\beta}'_{\rm GR}=-2$. This is shown in the right panel of Fig.~\ref{Fig:PressuresSmall}.

As we take stars of increasing compactness, the $1-2m/r$ term in the denominator of~\eqref{Eq:TOVbeta} becomes smaller, and the gradient of the pressure increases accordingly. Eventually, the TOV equation is dominated by the regime  $p'\propto p^2$, reaching a solution like~\eqref{eq:Psep}, which denotes an upper compactness bound. In the particular case depicted in Fig.~\ref{Fig:SmallStars}, the bound is given by $2M/R\approx 0.956$, appreciably above the Buchdahl limit $2M/R=8/9$.

Let us now consider fluid spheres placed inside the inner horizon. The term $-m\bm{\beta'}$ in the TOV equation~\eqref{Eq:TOVbeta} is now negative, and the pressure initially \textit{decreases} inwards. Since this term cannot further change sign for $r<R$ and $p(r)<0$, an inward-decreasing pressure cannot diverge nor change monotonicity, as it could only
reach a turning point if it acquires the value $p=-\rho_{0}$. However, this value cannot be intersected at a non-zero radius as it would be inconsistent with the TOV equation itself. Hence, stars placed inside the inner horizon can have a surface compactness lying between zero (arbitrarily small surface) and unity (surface arbitrarily close to the inner horizon from below). In the limit of compactness unity, these solutions approach a constant-pressure interior with $p=-\rho_{0}$ joined smoothly to its exterior, closely resembling the gravastar model~\cite{Mazur:2001fv,Visser:2003ge,Cattoen:2005he,Mottola:2010gp,Mazur:2015kia,Mottola:2023jxl,Bueno:2025tli} but without the need of thin shells or anisotropic pressures.
These profiles are shown in the left panel of Fig.~\ref{Fig:PressuresSmall}. Given that $p\in[-\rho_0,0]$ for all these solutions, the dominant energy condition (DEC) is thus satisfied while the strong energy condition (SEC) does not hold in general (see~\cite{Barcelo:2002bv,Curiel:2014zba,Martin-Moruno:2017exc,Martin-Moruno:2013wfa,Martin-Moruno:2015ena} for definitions); such behaviour is also observed in the analogous case with non-zero charge in general relativity~\cite{PoncedeLeon:2017usu}.

\begin{figure*}[htp] 
    \centering
    \subfloat{%
        \includegraphics[width=0.5\linewidth]{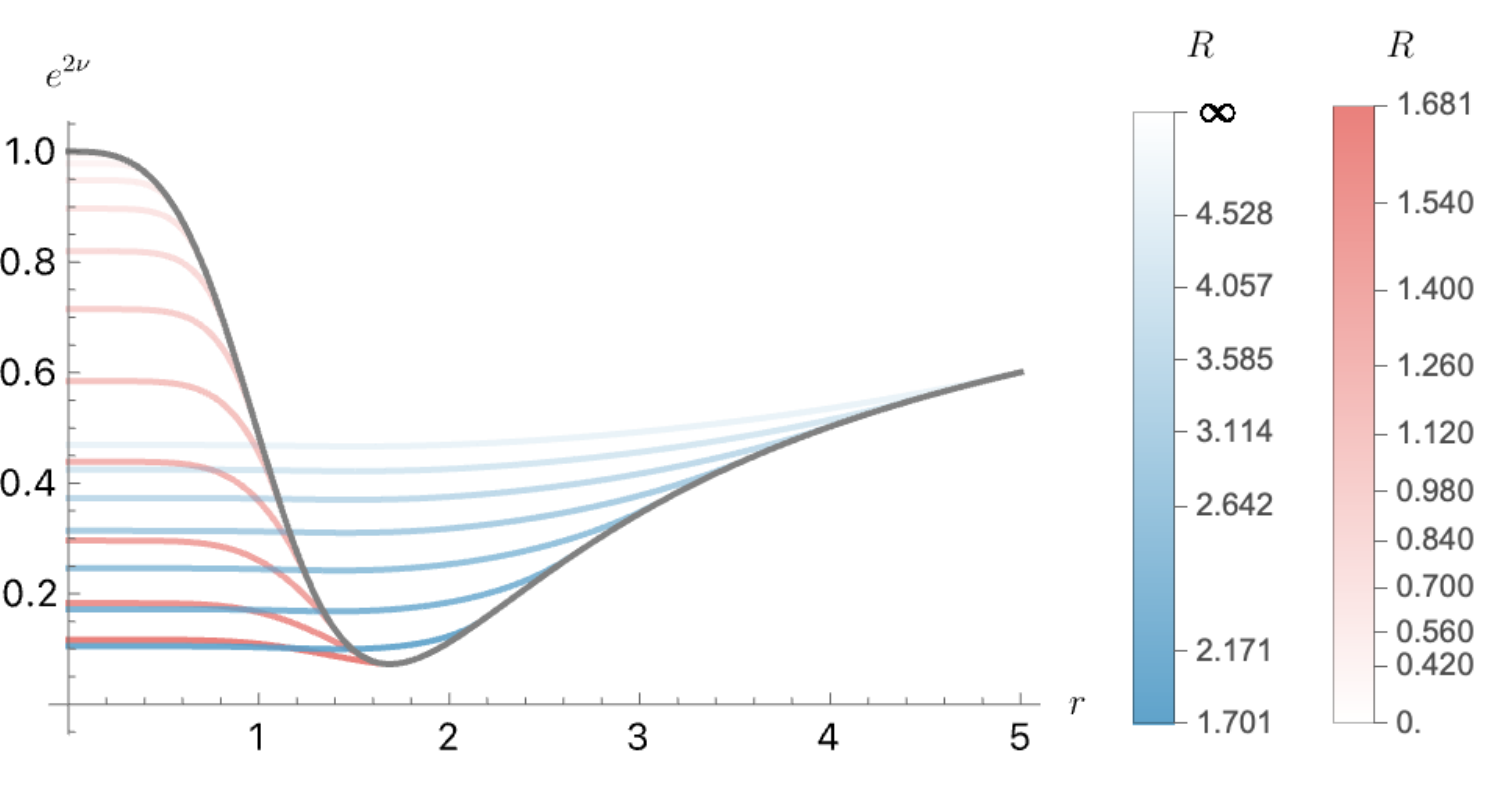}%
        \label{fig:a2}%
        }%
    \hfill%
    \subfloat{%
        \includegraphics[width=0.5\linewidth]{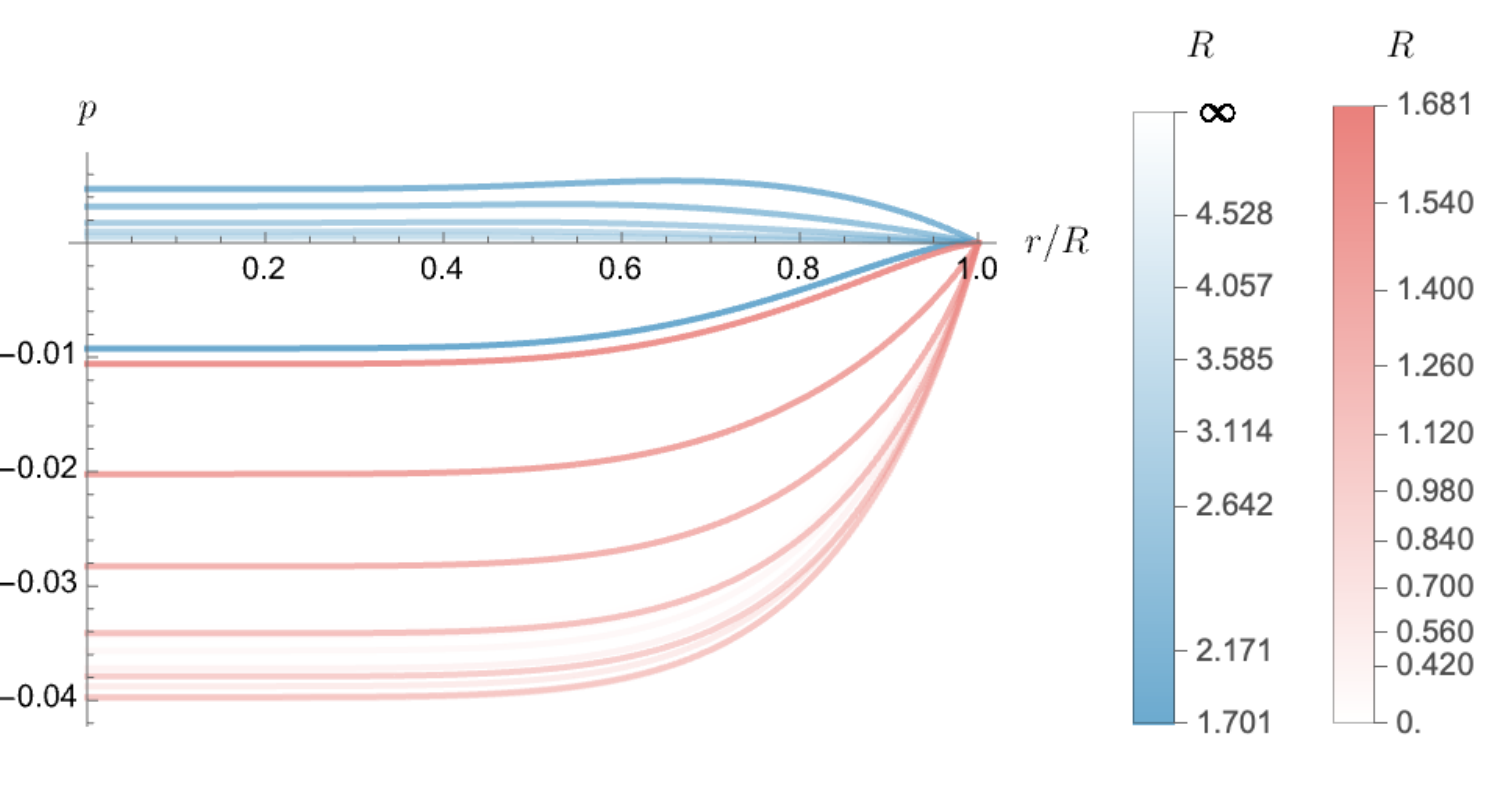}%
        \label{fig:b2}%
             }%
    \caption{Left panel: Redshift functions of stars placed outside (in blue) and inside (in red) the global minimum of the redshift. We have taken $M=1$ and $\ell=1.374$. When the vacuum solution (dark gray) does not exhibit horizons, regular fluid spheres can be placed anywhere in the spacetime. Their maximum compactness will be given by $1-e^{2\nu(r_{\text{min}})}$, and can tend to $1$ in the limit where $r_{\text{min}}\to0$. Right panel: Pressure profiles of these stars. As stars are compressed, their internal pressures eventually reach a maximum value. }
    \label{Fig:RedsPresExtr}
\end{figure*}

In summary, the maximum compactness bound for exterior stars depends on $\ell$, while interior stars can take any compactness from $0$ to $1$. What happens when $\ell$ is increased beyond the value that makes the horizons disappear in the vacuum solution? In that case, it is possible to construct fluid spheres of any radius. The resulting redshift and pressure profiles are shown in Fig.~\ref{Fig:RedsPresExtr}.

Stars placed at the region where $\nu'>0$ (for radius larger than the radius of the turning point for the redshift function) will now have finite pressure everywhere. This is because, if $\ell$ is large enough, the numerator of  ~\eqref{Eq:TOV2} can change sign due to the term $-m\bm{\beta}'$ becoming positive. Thus, pressure reaches a maximum and it is guaranteed to be regular everywhere. It is possible to place stars across the turning point of the redshift function and beyond. 

Note that there is a particular case where the compactness of stars placed at the exterior is equal to $1$. This case describes a vacuum solution where there is a single, degenerate horizon, that is, when the minimum in the redshift happens for zero redshift. In that case, we have, at the surface of the star, that
\begin{eqnarray}
    p'(R)=-\frac{\rho_{0}m\bm{\beta}'}{\left(r-2m\right)\bm{\beta}}=0,
\end{eqnarray}
and 
\begin{eqnarray}
    p''(R)=-\frac{\rho_{0}m\bm{\beta}''}{\left(r-2m\right)\bm{\beta}}<0,
\end{eqnarray}
so pressure starts from a maximum value of $0$ at the surface and decreases inwards towards negative values. As it is negative and decreasing, it's finite everywhere and the solution is regular. 

\begin{figure}
    \centering
    \includegraphics[width=0.7
    \linewidth]{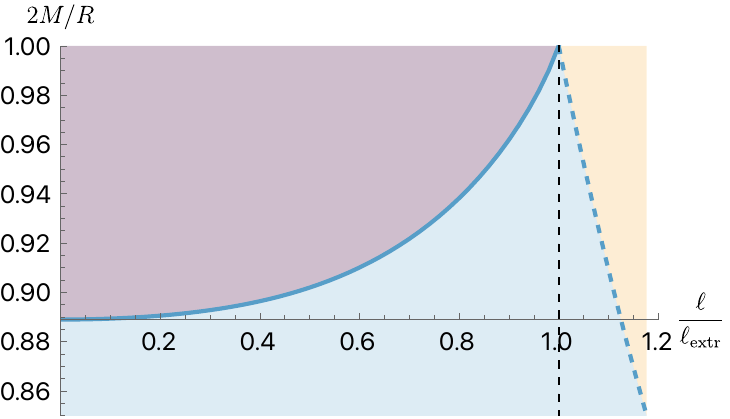}
    \caption{Space of stellar solutions in theories corrected by a lengthscale $\ell$. In blue and pink regions represent stars with finite and infinite pressures, respectively, while the yellow region is an inaccessible region of the parameter space with no stellar solutions. The maximum compactness of stars (blue line) grows with $\ell$ until the vacuum solution becomes horizonless, after which the maximum compactness of stars is bounded by and equal to the maximum compactness $2m(r)/r$ of the vacuum metric. }
    \label{Fig:Buchdahl}
\end{figure}

Figure~\ref{Fig:Buchdahl} shows the various regions of the space of stellar solutions and how they are modified by the value of $\ell$. These regions correspond to solutions with finite pressures in blue, infinite pressures in pink, and inaccessible (they do not match the vacuum solution smoothly) in yellow. For $\ell<\ell_{\rm{extr}}$, this limit grows monotonically as the outer horizon in the vacuum solution shrinks. The continuous blue line denotes the stellar solution with infinite central pressure, above which we find no stars with finite pressure. For $\ell=\ell_{\rm{extr}}$, the vacuum solution has a single, extremal horizon and the compactness of stars can reach values arbitrarily close to $2M/R=1$ both from inside and outside the horizon. For $\ell>\ell_{\rm{extr}}$, there is a maximum compactness bound in the vacuum solution given by the minimum of the redshift function. This compactness limit is not of infinite pressure, thus we have represented it by a dashed blue line.

\section{Conclusions}
\label{sec:Conclusions}

In this paper, we have constructed the most general form of the $D$-dimensional Tolman--Oppenheimer--Volkoff (TOV) equation obtained under the assumption that the spherically symmetric Einstein tensor is deformed into an identically conserved tensor containing up to second derivatives of the metric~\cite{Carballo-Rubio:2025ntd}. 

In this construction, fields defined in a lower-dimensional spacetime are interpreted as the building blocks of a higher-dimensional spherically symmetric spacetime in warped-product form~\cite{Kunstatter:2015vxa}. Additional conditions are thus required to guarantee the geodesic completeness of the reconstructed spherically symmetric spacetime~\cite{Carballo-Rubio:2025ntd}, which we have determined explicitly.

We have solved the TOV equation for constant density and specific families of deformations of general relativity, focusing mostly on the $D=4$ case for concreteness. The deformations considered belong to the ZK family, examples of which were considered more than a decade ago by Ziprick and Kunstatter for the study of the dynamical formation of regular black holes~\cite{Ziprick:2010vb}. However, in this paper we have focused on deformations that do not regularize vacuum solutions, motivated both by the conditions associated to geodesic completeness and the goal of considering situations that can serve to illustrate universal features. Our analysis is thus complementary to the recent analysis of stellar equilibrium in quasitopological gravities~\cite{Bueno:2025tli}, as the theories considered in both works are disjoint.

One of the universal features highlighted is the dependence of the Buchdahl limit on the new parameters characterizing the deformations with respect to general relativity. The family of deformations considered here is characterized by a single parameter $\ell$, and the dependence of the Buchdahl limit on the latter has been illustrated numerically. Due to the deformations resulting in a weakening of gravity, the resulting Buchdahl limit increases continuously with $\ell$ from its value in general relativity ($\ell=0$). 

Another universal feature, which has been also observed in quasitopological theories~\cite{Bueno:2025tli}, is the existence of solutions with an inner fluid core, the radius of which must not be larger than the corresponding radius of the inner horizon of the vacuum solutions associated with the deformation. These configurations are characterized by negative pressures (although satisfying the dominant energy condition). Interestingly, this is also the case in general relativity when charged fluids are considered, although in the current context this phenomenon arises due to the repulsive effects induced by the deformations of general relativity we are considering. The existence of solutions with inner fluid cores seem thus to rely only on the existence of inner horizons.

Going beyond the present equilibrium analysis to determine the stability properties of the configurations studied here is a clear direction to consider next, for which the master field equations provide an appropriate framework. It would be interesting to determine whether equilibrium solutions are generically stable for all possible deformations of general relativity (e.g., both the ZK and KMT families), or whether differences arise there. The stability of solutions with inner fluid cores is also an open question due to the well-known instability of inner horizons~\cite{Carballo-Rubio:2024dca}. It is likely that said instability will also show up in an unstable behavior of the inner fluid core itself, which could be revealed by studying fluid perturbations in a similar way as in general relativity.

\appendix

\section{Junction conditions}

\label{app:junc}

Here we sketch the derivation of the junction conditions for the field equations~\eqref{eq:feqs} in static situations, using a proper distance radial coordinate $x^1=z$ in which $q_{11}=1$, so that the 2-dimensional line element in Eq.~\eqref{eq:2gstaticchi} becomes
\begin{equation}\label{eq:linelz}
q_{ab}(x)\text{d}x^a\text{d}x^b=-q_{00}(z)\text{d}t^2+\text{d}z^2.
\end{equation}
In these coordinates, possible distributional components arise from $q_{00}(z)$ and $r(z)$ at the surface of the star $z_R=z(R)$~\cite{Israel:1966rt}. We will also restrict the discussion here to the ZK family of theories in four dimensions.

Using these special coordinates, the first junction condition implies the continuity of the metric functions, $[q_{00}]=q_{00}^+-q_{00}^-=0$ and $[r]=r^+-r^-=0$. On the other hand, the derivatives along $z$ can be discontinuous, and thus $[\partial_z q_{00}]=\partial_z q_{00}^+-\partial_z q_{00}^-$ and $[\partial_z r]=\partial_z r^+-\partial_z r^-$ can be non-zero in principle. These discontinuities have a clear physical interpretation in terms of a (singular) surface stress-energy tensor. Hence, the tensor $\mathscr{G}_{\mu\nu}$ and $\mathscr{F}$ defined in Eq.~\eqref{eq:mfeqs} has a distributional component $8\pi\delta\left(z-z_R\right)\mathscr{S}_{\mu\nu}$~\cite{Poisson:1989zz,Mars:1993mj}, which can be calculated explicitly using Eqs.~(\ref{eq:geneqs1}-\ref{eq:geneqs2}):
\begin{align}\label{eq:distten}
\mathscr{S}_{00}&=\frac{1}{8\pi r^{2}}[\partial_zr]\bm{\beta},\quad \mathscr{S}_{01}=\mathscr{S}_{11}=0,\nonumber\\
\mathscr{S}_{ij}&=-\frac{r}{32\pi}\gamma_{ij}\left\{-\frac{[\partial_zq_{00}]}{q_{00}}\bm{\beta}+2[\partial_zr]\partial_r\bm{\beta}\right\}.
\end{align}
These expressions reduce to the ones in general relativity (see, e.g.,~\cite{Visser:1995cc}) for $\bm{\beta}=\bm{\beta}_{\rm GR}=-2r$:
\begin{align}
\mathscr{S}_{00}&=\-\frac{[\partial_zr]}{4\pi r},\quad \mathscr{S}_{01}=\mathscr{S}_{11}=0,\nonumber\\
\mathscr{S}_{ij}&=\gamma_{ij}\left\{\frac{[\partial_zr]r}{8\pi}-\frac{[\partial_zq_{00}]r^2}{16\pi q_{00}}\right\}.
\end{align}
All these components are zero at the surface of a constant-density star, since $[\partial_zq_{00}]=[\partial_zr]=0$ at the boundary. Furthermore, we observe there is no choice of $\bm{\beta}$ function that can make all components of this distributional stress-energy tensor vanish. The only effect of $\bm{\beta}$ its to modify the coefficients in the components of the stress-energy tensor of the shell.

The following relations are useful to make the connection between the discussion here and the coordinates used in Sec.~\ref{Sec:SteEqu}. Direct comparison between Eqs.~\eqref{eq:2gstaticchi} and~\eqref{eq:linelz} shows that $e^{2\nu}=q_{00}$ and $\chi=\left(\partial_zr\right)^2$, so that
\begin{eqnarray}
    \nu'=\frac{\partial_zq_{00}(\partial_zr)^{-1}}{2q_{00}},\quad \partial_zr=\sqrt{1-\frac{2m(r)}{r}}.
\end{eqnarray}
The condition $[\partial_zr]=r^+-r^-=0$ implies then $[m]=0$ and, together with $[\partial_zq_{00}]=\partial_zq_{00}^+-\partial_zq_{00}^-=0$,
\begin{equation}
[\nu']=\frac{1}{2q_{00}}\left(\frac{\partial_{z}q_{00}^{+}}{\partial_{z}r^{+}}-\frac{\partial_{z}q_{00}^{-}}{\partial_{z}r^{-
    }}\right)=\frac{\partial_zq_{00}^+\partial_zr^--\partial_zq_{00}^-\partial_zr^+}{2q_{00}\partial_zr^+\partial_zr^-}=0.    
\end{equation}
We thus obtain the conditions in Eq.~\eqref{Eq:JunctionR}.

\acknowledgments

R.C.R. acknowledges support by the Spanish Government
through the Grant No. PID2023-
149018NB-C43 funded by MCIN/AEI/10.13039/501100011033, and the Severo Ochoa grant CEX2021-001131-S funded by MCIN/AEI/ 10.13039/501100011033.

\bibliographystyle{utphys}

\bibliography{main_refs}

\end{document}